\documentclass[aps,prl,twocolumn,superscriptaddress]{revtex4-2}
\usepackage{graphicx}
\usepackage{color,bm}
\usepackage[colorlinks,bookmarks=false,citecolor=blue, urlcolor=blue]{hyperref}
\usepackage{amsmath,amssymb}
\usepackage{mathrsfs}

\renewcommand{\vec}[1]{\bm{#1}}

\newcommand{\beginsupplement}{        \setcounter{table}{0}
	\renewcommand{\thetable}{S\arabic{table}}        \setcounter{figure}{0}
	\renewcommand{\thefigure}{S\arabic{figure}}               
}

\begin{document}

\title{Screw dislocations in cubic chiral magnets}

\author{Maria Azhar}
\affiliation{Institut f\"ur Theoretische Festk\"orperphysik, Karlsruhe Institute of Technology, 76131 Karlsruhe, Germany}

\author{Volodymyr P. Kravchuk}
\affiliation{Institut f\"ur Theoretische Festk\"orperphysik, Karlsruhe Institute of Technology, 76131 Karlsruhe, Germany}
\affiliation{Bogolyubov Institute for Theoretical Physics of National Academy of Sciences of Ukraine, 03680 Kyiv, Ukraine}

\author{Markus Garst}
\affiliation{Institut f\"ur Theoretische Festk\"orperphysik, Karlsruhe Institute of Technology, 76131 Karlsruhe, Germany}
\affiliation{Institute for Quantum Materials and Technology, Karlsruhe Institute of Technology, 76131 Karlsruhe, Germany}

\date{\today}

\begin{abstract}
Helimagnets realize an effective lamellar ordering that supports disclination and dislocation defects. Here, we investigate the micromagnetic structure of screw dislocation lines in cubic chiral magnets using analytical and numerical methods. The far field of these dislocations is universal and classified by an integer strength $\nu$ that characterizes the winding of magnetic moments. We demonstrate that a rich variety of dislocation-core structures can be realized even for the same strength $\nu$. In particular, the magnetization at the core can be either smooth or singular. We present a specific example with $\nu = 1$ for which the core is composed of a chain of singular Bloch points. In general, screw dislocations carry a non-integer but finite skyrmion charge so that they can be efficiently manipulated by spin currents.
\end{abstract}

%

\maketitle

\begin{figure}
\includegraphics[width=\columnwidth]{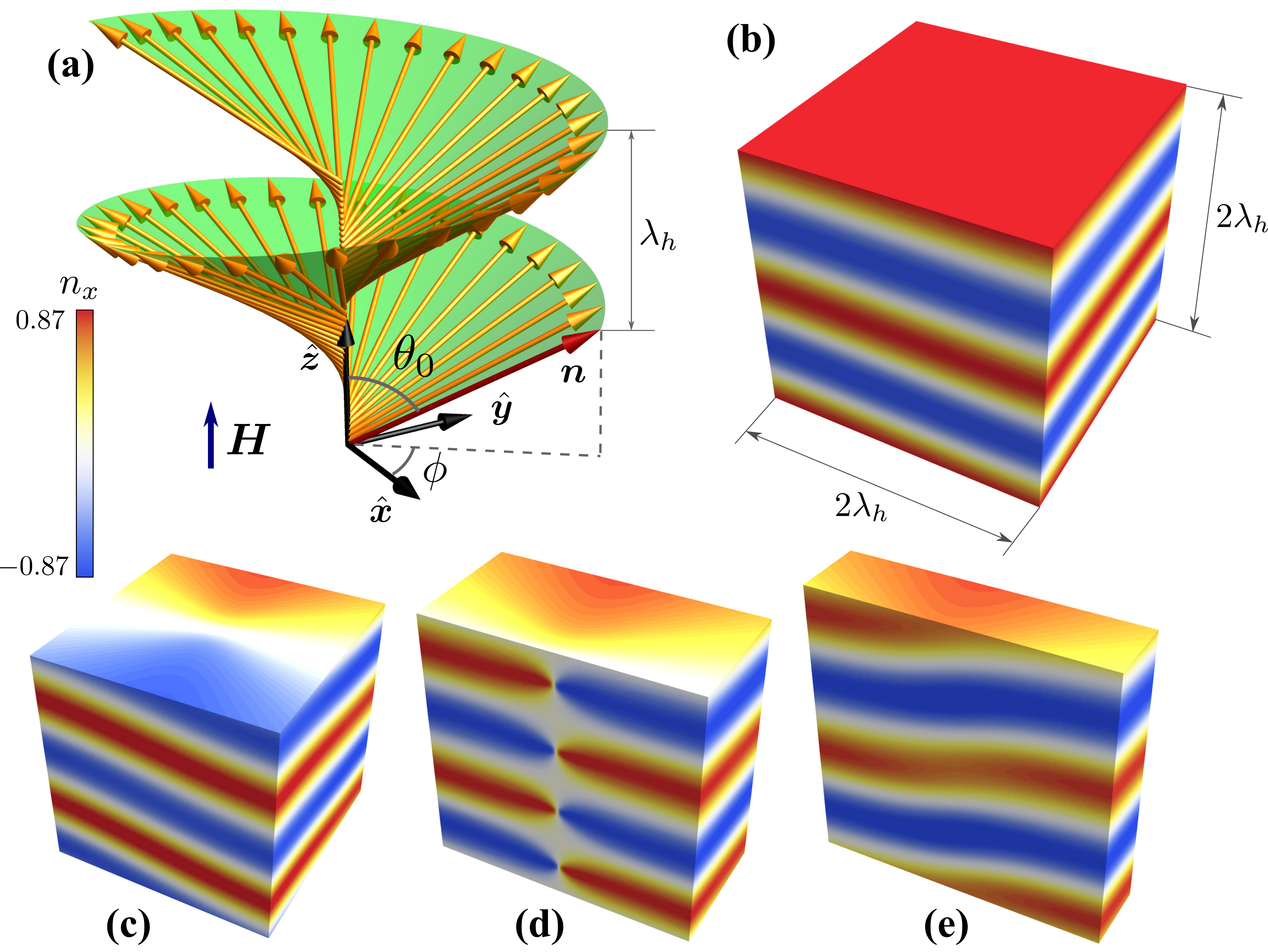}
\caption{(a) Conical magnetic helix with pitch $\lambda_h$ and cone angle $\theta_0$ enclosed by the magnetic moments and the direction of the applied magnetic field $\boldsymbol H$. (b) Helimagnetic order is characterized by equidistant isosurfaces, the hallmark of lamellar order, where, e.g., the x-component of magnetization, $n_x$, assumes a particular value. (c)-(e) Example of a screw dislocation with strength $\nu=1$ illustrated by different vertical cross-sections. The core of this specific example contains a chain of magnetic Bloch points.
}
\label{fig:conical}
\end{figure}

{\it Introduction --}
The Dyzaloshinskii-Moriya interaction (DMI) in cubic chiral magnets  like MnSi, FeGe, or Cu$_2$OSeO$_3$ favours helimagnetic long-range order in a large region of the phase diagram \cite{BakJensen1980,Nakanishi1980,Ishikawa1976,Lebech1989,Uchida2006,Adams2012,Seki2012}.  A finite field $\boldsymbol{H}$ aligns the helix axis and, in addition, tilts the magnetic moments towards the field direction giving rise to a conical magnetic helix, see Fig.~\ref{fig:conical}(a). This helimagnetic ordering realizes a one-dimensional periodic texture that shares many similarities with other emerging lamellar structures found, e.g., in various soft matter systems \cite{Chaikin00,P.G.deGennes95,Kleman04}. 

In particular, in the limit of weak spin-orbit coupling (SOC) the phase transition from the paramagnetic to the helimagnetic phase at $\boldsymbol{H} = 0$ is a fluctuation-driven first-order transition similar to the ones in certain cholesteric liquid crystals or diblock co-polymers \cite{Janoschek2013,Buhrandt2013,Zivkovic2014}. The correlation length above the critical temperature $T_c$ possesses a temperature dependence that is well described by weak crystallization theory \cite{Brazovskii1987}. This indicates that the paramagnetic regime just above $T_c$ is characterized by strong correlations that are maintained by the large density of states of paramagnons \cite{Kindervater2019}. It is still an important open issue whether these pronounced magnetic correlations are also at the origin of the non-Fermi liquid behavior observed in MnSi and FeGe upon suppressing the critical temperature towards zero with pressure \cite{Pfleiderer2001,Pfleiderer2007,Pedrazzini2007,Ritz2013,Schmalian2004}. 

In this context, the question arises as to whether this intriguing paramagnetic-helimagnetic phase transition can also be understood from the dual perspective as a defect-mediated melting transition \cite{Toner1982,Grinstein1986,Zhai2021}. A prerequisite to address this question is an understanding of the elementary defects of helimagnetic order. It is well-known that defects of  lamellar structures in general consist of disclinations and dislocations, which are line defects in case the lamellae are embedded in three dimensional space \cite{Chaikin00}. For helimagnets, such defects were discussed on a phenomenological level by Kl{\'{e}}man \cite{Kleman70}. Recently, it was shown both theoretically and experimentally that domain walls of helimagnetic order might consist of an arrangement of disclinations and edge dislocations \cite{Li2012,Nattermann2018,Schoenherr18} very similar to domain walls in cholesteric liquid crystals \cite{Bouligand1983}. 
Moreover, at helimagnetic twist grain boundaries screw dislocations are expected to occur \cite{Chaikin00,Martin2017}. It was also demonstrated that the motion of edge dislocations is an important relaxation process for disordered helimagnets possibly accounting for the large relaxation times observed experimentally \cite{Bauer2017,Milde2020,Dussaux16,schoenherr2021}. 

However, there exist additional line excitations within the conical helix phase that are distinct from dislocations and disclinations. In particular, chiral magnets are famously known to host skyrmions, i.e., topological two-dimensional magnetic textures \cite{Bogdanov1994}. In bulk magnets, the skyrmion textures extend along the third direction forming skyrmion strings that either condense into a lattice or exist as metastable excitations of the field-polarized phase \cite{Muhlbauer2009,Yu2010,Seki2012}, for a recent review see Ref.~\cite{Back_2020}. It has been demonstrated in \cite{Leonov16a} that such metastable skyrmion configurations also persist within the conical helix phase upon decreasing the magnetic field below the critical field $H_{c2}$. Such skyrmion strings attract each other and can form clusters or even networks \cite{Du2018,Sohn2019,Leonov21}. Nevertheless, these skyrmion strings within the conical helix phase 
possess an exponentially decaying far field and,  in contrast to dislocation lines, are characterized by a vanishing Burgers vector. In addition, the conical helix can also support localized large-amplitude excitations like bound pairs of hedghog defects, i.e., Bloch points \cite{Mueller2020} and even Hopfions \cite{Voinescu2020}. 
 
In the present work, we theoretically investigate in detail screw dislocations in cubic chiral magnets. The helimagnetic order defines equally spaced isosurfaces where, e.g., the $x$-component of the magnetization assumes the same value, see Fig.~\ref{fig:conical}(b). The deviation of isosurfaces from their equilibrium configuration is described by the displacement field $\boldsymbol{u}$ \footnote{The displacement field $\vec{u}$ describes small deviations of the fronts of constant phases $\phi(\vec{r})=\text{const}$ of the helimagnetic order where $\phi(\vec{r})=2\pi (\vec{r}+\vec{u}) \hat{\vec{z}}/\lambda_h$.}. The integral along a loop enclosing a dislocation line, $\oint d\boldsymbol{u} = \boldsymbol{b}$, is finite and given by the Burgers vector $\boldsymbol{b}$ indicating that $\boldsymbol{u}$ is singular at the dislocation core. For a screw dislocation, $\boldsymbol{b} = \lambda_h \nu \hat{\vec{z}}$ is aligned with the helix axis $\hat{\vec{z}}$ and its size is an integer multiple of the helix pitch $\lambda_h$, where $\nu \in \mathbb{Z}\setminus\{0\}$ characterizes the strength of the screw dislocation (sd$_\nu$). 

Using analytical arguments and numerical simulations we determine the micromagnetic structure of screw dislocations. In the limit of small SOC, when the influence of magnetocrystalline anisotropies is negligible, we find that they possess in the far field the expected universal form of lamellar structures with a displacement vector $\boldsymbol{u} = u_z(x,y) \hat{\vec{z}}$ where \cite{Chaikin00,P.G.deGennes95,Kleman04}
\begin{align} \label{DisplacementField}
u_z(x,y) =  \frac{\lambda_h}{2\pi} \nu  \chi.
\end{align}
Here, $\chi$ is the polar angle of cylindrical real-space coordinates $(\rho, \chi, z)$.
Moreover, we show that the magnetization texture at the core of screw dislocations can either be smooth or might contain Bloch points, which we illustrate explicitly for screw dislocations with $|\nu| = 1$. 

{\it Theory of cubic chiral magnets --}
The magnetic energy functional $E = \int d\vec r \mathcal{E}$ of cubic chiral magnets possesses a density that reads in leading order in SOC
\begin{align}
\label{eq:H}
\mathcal{E} = A (\partial_i \vec{n})^2 + D \vec{n} (\vec{\nabla} \times \vec{n}) - M_s \mu_0 H  n_z.
\end{align}
Here, $\vec n$ is a unit vector specifying the orientation of the local magnetization and the magnetic field is applied in the $z$-direction, $\boldsymbol H = H \hat{\vec{z}}$; $A$ is the exchange constant, $D > 0$ is the DMI assuming a right-handed chiral magnetic system, $M_s$ is the saturation magnetization and $\mu_0$ is the magnetic constant. Importantly, for zero field $H= 0$ this density is isotropic with respect to a combined rotation of spin and real space. This rotational symmetry is explicitly broken by magnetocrystalline anisotropies that are however weak in the limit of small SOC and will be mostly neglected in the following. We also neglect for simplicity the magnetic dipolar interaction.

It is convenient to consider the representation of $\vec n = (\sin\theta \cos \phi, \sin\theta \sin \phi, \cos \theta)$ in terms of polar angle $\theta$ and azimuthal angle $\phi$. For large fields $H > H_{c2} = D^2/(2A \mu_0 M_s)$ the ground state of Eq.~\eqref{eq:H} is field-polarized with $\theta = 0$. The ground state for small fields $0 \leq H < H_{c2}$ is the conical helix with a position dependent polar angle $\phi = 2\pi z/\lambda_h$ where the helix pitch $\lambda_h = 4\pi A/D$, and the cone angle $\theta = \theta_0$ with  $\cos \theta_0 = H/H_{c2}$. 

{\it Far field of screw dislocations --}
Performing an asymptotic analysis of the Euler-Lagrange equations of Eq.~\eqref{eq:H}, for details see the supplementary material \cite{SI}, we find that the conical helix supports screw dislocations with an asymptotic behavior for large distances from their core $\rho \to \infty$, 
\begin{align}
\label{ThetaFarField}
\theta &= \theta_0 + \frac{2 \nu \sin^2 \theta_0}{1+\sin^2 \theta_0} \frac{\lambda_h}{2\pi \rho} \sin \big( (\nu-1) \chi + \frac{2\pi z}{\lambda_h}\big)
+ \mathcal{O}(\rho^{-2}), \\
\label{PhiFarField}
\phi &= \frac{2\pi}{\lambda_z} \Big(z + u_z(x,y) \Big) +  \mathcal{O}(\rho^{-2}),
\end{align}
where $u_z$ is the universal displacement field of Eq.~\eqref{DisplacementField}. The dependence of $\phi$ indicates that the $(x,y)$-components of the magnetization form a vortex within each plane perpendicular to the applied field, i.e., for each value of $z$, and the winding number is just given by the disclination strength $\nu$. The structure of the vortex changes from plane to plane as a function of $z$ due to the linear dependence of $\phi$ on $z$, see Fig.~\ref{fig:conical}.

The far-field configuration allows to determine the topological skyrmion charge for a screw dislocation within each $z$-plane, $N_{\rm top}(z) = \int d x dy \rho_{\rm top}$ where $\rho_{\rm top} = \frac{1}{4\pi} \vec{n}(\partial_x \vec{n} \times \partial_y \vec{n})$. Note that, in contrast to skyrmions, here $N_{\rm top}(z)$ is not an integer.
At infinity $\rho \to \infty$, the magnetization $\vec{n}$ encircles as a function of real-space angle $\chi$ $\nu$-times the $\hat{\vec{z}}$ axis with a fixed polar angle $\theta_0$.
Assuming that the magnetization texture is smooth in a given $z$-plane, we obtain for the charge
\begin{align}
N_{\rm top}(z) = \frac{\nu}{2} \left(1-h \right) + n(z),
\end{align}
with the reduced field $h = H/H_{c2}$, and $n(z) \in \mathbb{Z}$ is an integer that depends on the magnetization at the core of the screw dislocation. As we will see below, in case that $n(z)$ varies with $z$ the core is singular and contains Bloch points.

\begin{figure}
\includegraphics[width=\columnwidth]{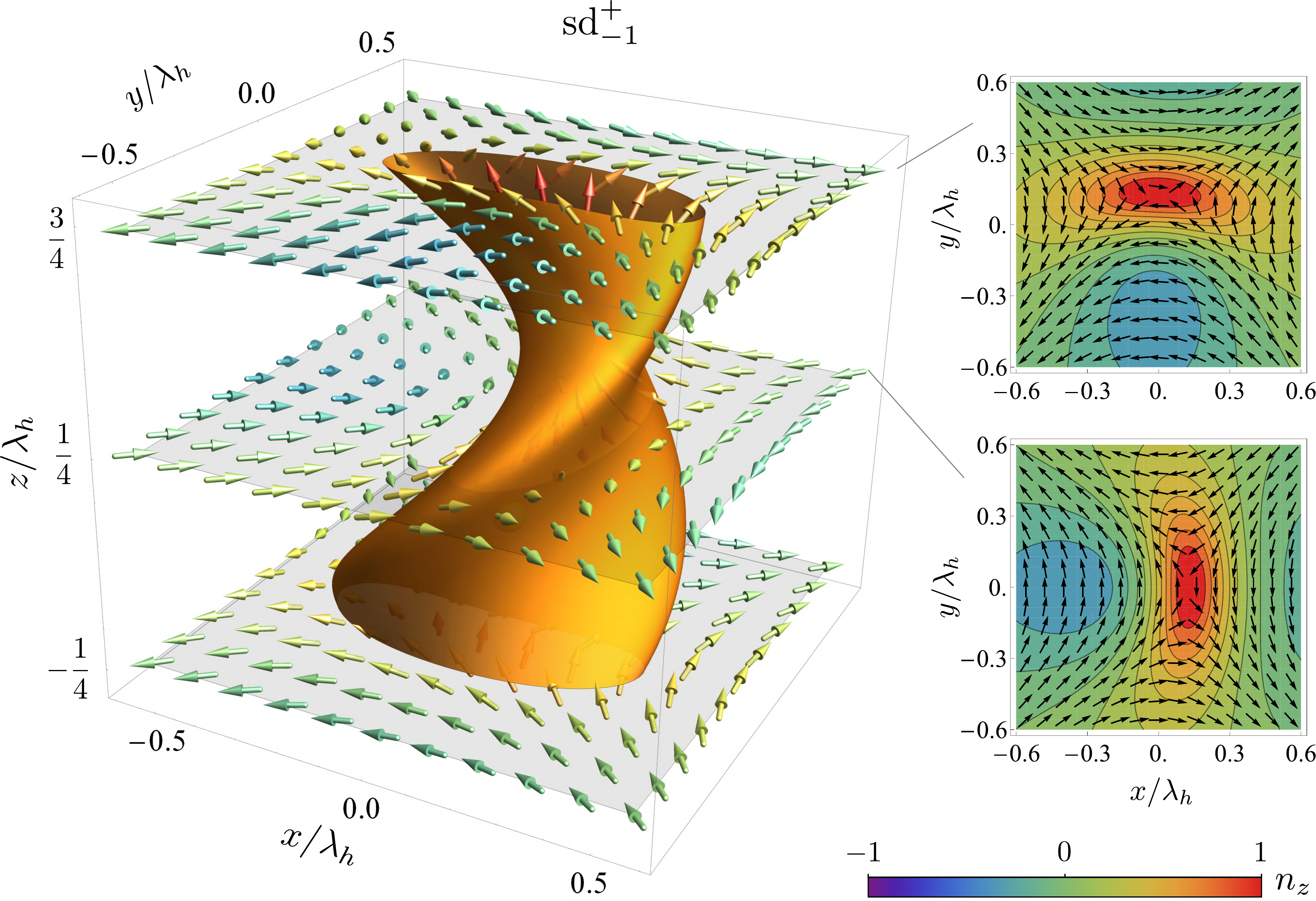}
\caption{
Core of the screw dislocation sd$^+_{-1}$ with strength $\nu = -1$ obtained by micromagnetic simulations for $H=0$. The magnetic moments represented by arrows form within a given $z$-plane an antivortex-like structure. The orange isosurface in the main panel is defined by $n_z = 1/2$. The magnetization at the core is preferentially pointing in $\hat{\vec{z}}$ direction; the configuration with opposite core magnetization, sd$^-_{-1}$, is degenerate at $H=0$ (not shown). Note that the periodicity of the core structure along $z$ is characterized by a wavelength $2\lambda_h$.
}		
\label{AVd}
\end{figure}

Plugging the asymptotics of Eqs.~\eqref{ThetaFarField} and \eqref{PhiFarField} into Eq.~\eqref{eq:H} we obtain for the energy of a screw dislocation line per length
\begin{align} \label{EnergySD}
\varepsilon_{\rm sd} = \varepsilon^{\rm core}_{\rm sd} + A (2 \pi  \nu)^2  \frac{h^2 (1-h^2)}{2-h^2} \log \frac{R}{\rho_{\rm core}},
\end{align}
where $h = H/H_{c2}$. The length scale $R$ specifies the extension of the system in radial direction, and 
$\rho_{\rm core}$ is the linear size of the dislocation core with the associated core energy $\varepsilon^{\rm core}_{\rm sd}$. The far-field tail of the screw dislocation gives rise to a contribution to the energy that in general diverges logarithmically with the radial system size $R$. This logarithmic contribution vanishes at the transition $H = H_{c2}$ to the  field-polarized phase because the dislocation ceases to be defined when the cone angle vanishes, $\theta_0 =0$. Moreover, it also vanishes in zero field $H=0$ due to the rotational symmetry of Eq.~\eqref{eq:H}. It is well-known that lamellar structures emerging in an isotropic environment are characterized by particularly soft small-amplitude, i.e., phonon excitations that possess the Landau-Peierls form \cite{Chaikin00}. In this case the contribution  to the energy from the far field of screw dislocations, which can be captured in terms of a static phonon field, vanishes \cite{Chaikin00,P.G.deGennes95,Kleman04,Santangelo06} in agreement with Eq.~\eqref{EnergySD}. Technically, this is here due to a cancellation of exchange and DMI energies. The prefactor of the logarithmic contribution in Eq.~\eqref{EnergySD} remains finite however at $H=0$ if magnetocrystalline anisotropies are taken into account that explicitly break the continuous rotational symmetry of the theory \cite{SI}. 

\begin{figure}
\includegraphics[width=0.9\columnwidth]{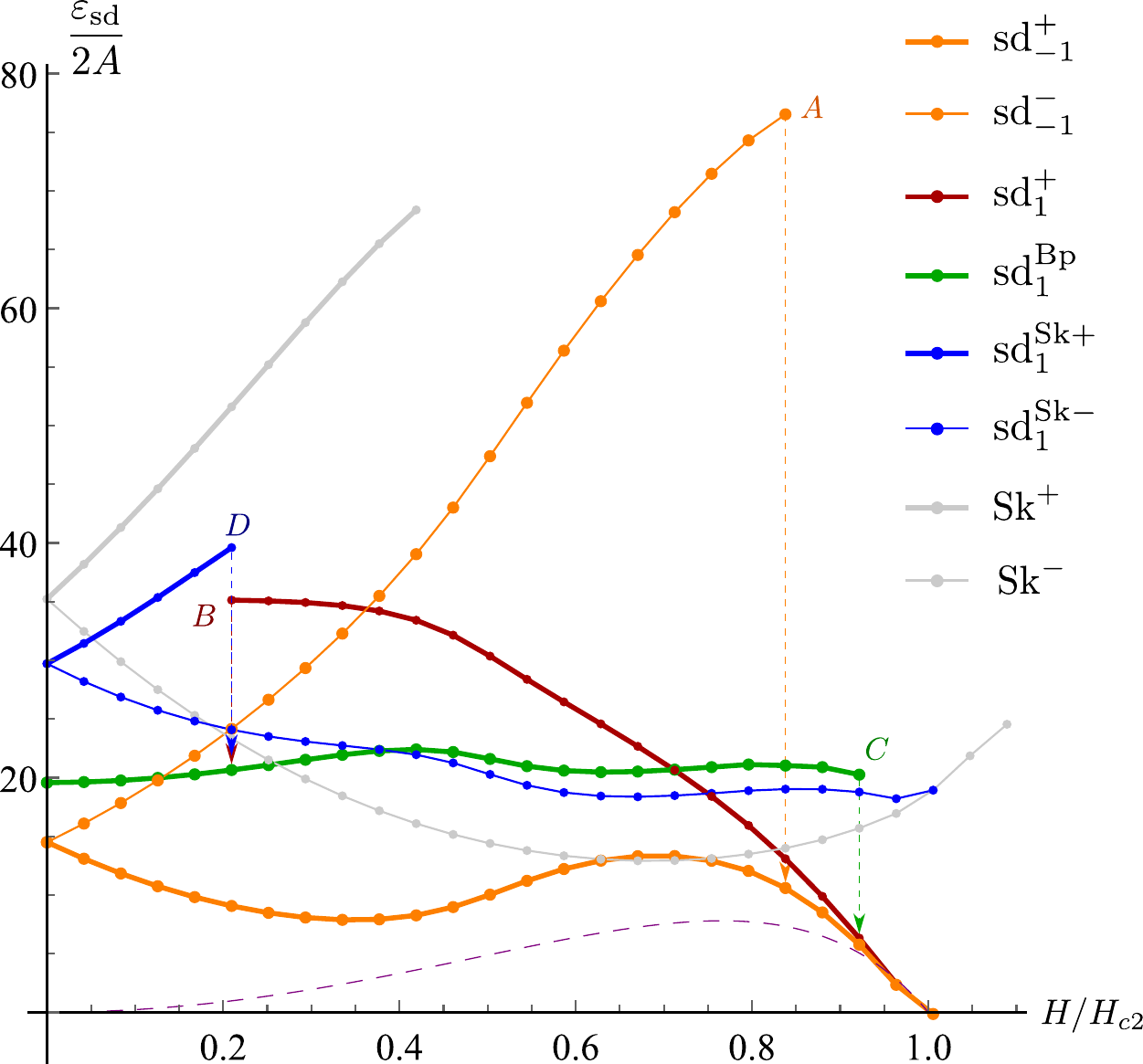}
\caption{Energy per length of the various screw dislocation lines shown in Figs.~\ref{AVd} and \ref{Vd}, as a function of $H$
with a comparison to the skyrmion string energy (gray lines) whose core magnetization is aligned (Sk$^+$) or anti-aligned (Sk$^-$) with the field. The energies were obtained by micromagnetic simulations for a cylinder-shaped system with radius $R = 5 \lambda_h$. Dashed line shows the logarithmic contribution of the far field in Eq.~\eqref{EnergySD} assuming $\rho_{\text{core}}=\lambda_h/2$ for illustration (for simplicity we assume that $\rho_{\text{core}}$ does not depend on $H$). The energy of dislocations sd$^+_{-1}$ and sd$^+_{1}$ vanishes at $H_{c2}$ where they can be identified as vortex lines of the XY-order parameter. The energy of sd$^{{\rm Sk}-}_{1}$ merges at $H_{c2}$ with that of the skyrmion string Sk$^-$. Points $A$-$D$ mark field values where the corresponding configurations became unstable in numerical simulations due to the lattice discreteness.
}\label{fig:energies}
\end{figure}

{\it Core structure of screw dislocations --}
Having established the far field of screw dislocations we now turn to the discussion of their core structures. We employ micromagnetic simulations \cite{SI} in order to determine the core and its energy. First, we focus on the case with strength $\nu = -1$, see Fig.~\ref{AVd}. The magnetization can here be continuously extrapolated from the far field towards the core resulting in a smooth texture. There exist two energetically degenerate configurations at zero field where the core magnetization is either aligned or anti-aligned with the field, respectively denoted by sd$^+_{-1}$ and sd$^-_{-1}$ in the following. In order to decrease DMI energy, the core deforms elliptically such that it mimics a small Bloch-domain wall with preferred chirality. In addition, we found that close to zero field the ellipse is further deformed into a banana-shape structure, see cross-sections in Fig.~\ref{AVd}, that leads to a periodicity of the core along the $z$-axis with an enhanced wavelength $2\lambda_h$. For finite field $H$, the configuration with the aligned core magnetization sd$^+_{-1}$ is energetically favoured, see Fig.~\ref{fig:energies}. The precise $H$-dependence of the dislocation energy  depends on the system size $R$ but it vanishes for $H \to H_{c2}$.

\begin{figure*}
\includegraphics[width=\textwidth]{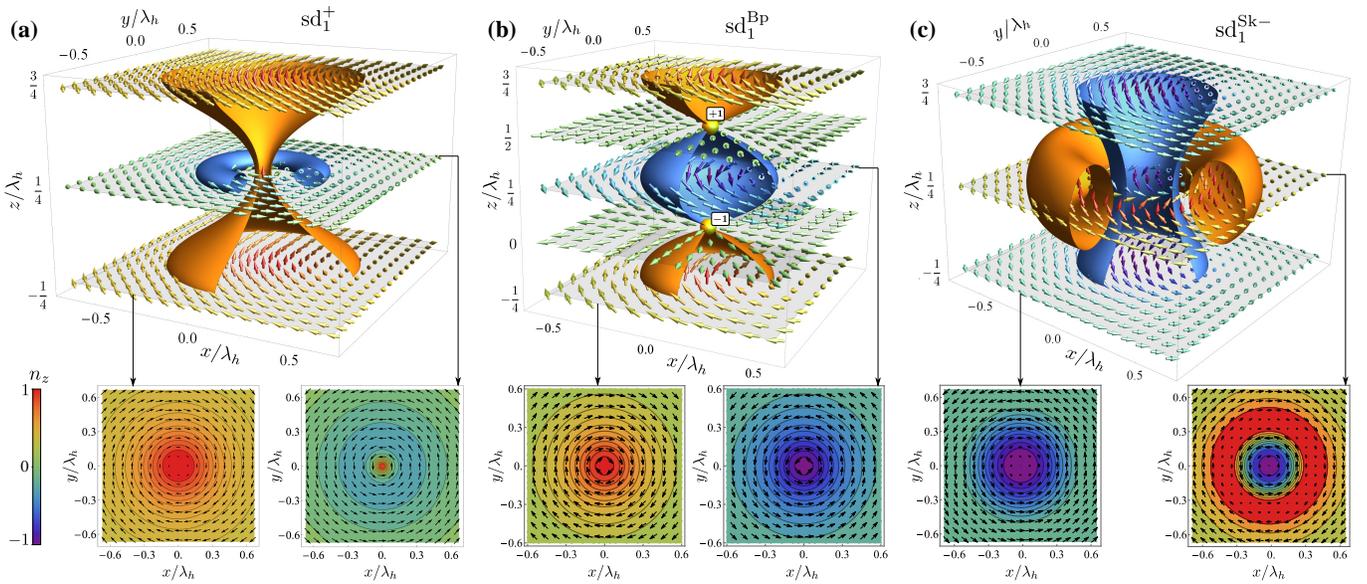}
\caption{
Core of various screw dislocations with strength $\nu = 1$ obtained by micromagnetic simulations. Orange and blue isosurfaces are respectively defined by $n_z = 1/2$ and $n_z = -1/2$ except in panel (a) where $n_z = 1/2$ and $n_z = -1/4$. (a) Dislocation sd$^{+}_{1}$ with core magnetization aligned with the field for $H = 0.21 H_{c2}$. (b) Dislocation sd$^{{\rm Bp}}_{1}$ for $H=0$ with an alternating core magnetization separated by Bloch points (yellow spheres) with alternating topological charges $\pm 1$. (c) Dislocation sd$^{{\rm Sk}-}_{1}$ at $H=0$ with an anti-aligned core magnetization; it smoothly connects to a skyrmion string configuration for $H \to H_{c2}$.
}
\label{Vd}
\end{figure*}

There exists in fact a screw dislocation for each strength $\nu$ whose energy vanishes at $H_{c2}$. The phase transition at $H_{c2}$ corresponds to a magnon condensation \cite{Giamarchi2008} that is in the XY-universality class. Employing a standard Holstein-Primakoff expansion around the field-polarized state $\vec{n} = \hat{\vec{z}} + (\psi\, e^{-i 2\pi z/\lambda_h} (\hat{\vec{x}} + i \hat{\vec{y}}) + c.c.) + \mathcal{O}(|\psi|^2)$  for $H > 0$ the complex spin wave function $\psi$ can be identified with the corresponding XY-order parameter. When it condenses for $H \lesssim H_{c2}$ with non-zero constant $\psi$, long-range conical order emerges. The $U(1)$ symmetry of the complex wave function $\psi$ then also supports vortex line solutions given by $\psi \sim e^{- i \nu \chi}$ with an amplitude vanishing at the core. The size of the vortex core is determined by 
the correlation length of the condensate, $\xi \sim 1/\sqrt{1-H/H_{c2}}$ \cite{Ivanov95b}, and is thus independent of $z$.
This implies a cylindrical core of the vortex line close to $H_{c2}$.
Hence, these vortex lines can be identified with a special type of screw dislocations of the helimagnetic order that possess a smooth magnetization at the core that is aligned with the applied field.

Let us consider the corresponding solution with strength $\nu = 1$ denoted by sd$^{+}_{1}$ in Fig.~\ref{Vd}(a). Here, the vortex structure of the $(x,y)$-components of magnetization in the far field continuously alternate from a divergenceless to a rotationless configuration as a function of $z$ with a topological charge $N_{\rm top} =\frac{1}{2} (1-H/H_{c2})$ that is independent of $z$. It is instructive to focus on the planes $z = \frac12(m \pm \frac{1}{2}) \lambda_h$ in Eq.~\eqref{PhiFarField} with $m \in \mathbb{Z}$ where a divergenceless configuration is realized. As the core is approached, the magnetization smoothly rotates in a right-handed or left-handed manner until it is aligned with the field at the center, see lower panels in Fig.~\ref{Vd}(a). Whereas the former texture is  favoured by the DMI, the latter is disfavoured. As a consequence, the core, that is cylindrical close to $H_{c2}$, becomes undulated along the $z$-axis for smaller fields with a contraction on the $z$-planes housing the disfavoured textures. The energetic cost of the latter also leads to an increase of the dislocation energy for decreasing $H$. In the simulations we found that this particular screw dislocation structure cannot be maintained for lowest magnetic fields. Instead, a first order transition to a different structure occurs at point $B$ in Fig.~\ref{fig:energies}, to which we turn next.

The energy cost of the disfavoured configurations shown in the lower right panel of Fig.~\ref{Vd}(a) can be avoided by switching the core magnetization within these planes. An alternating core magnetization along the $z$-axis is indeed characteristic for the screw dislocation sd$^{\rm Bp}_1$ shown in Fig.~\ref{Vd}(b). At zero field, it realizes a Bloch-like meron structure on the planes $z = \frac12(m \pm \frac{1}{2}) \lambda_h$ that are both favoured by the DMI. These merons possess alternating skyrmion charges $N_{\rm top} = \pm \frac{1}{2}$ that implies the presence of Bloch points with alternating topological charges $\pm 1$ \footnote{The topological charge $c$ of a Bloch point is defined by $c = \int_V \mathrm{d}\vec{r} \vec{\nabla} \cdot\vec \Omega$ where $V$ is a small volume around the Bloch point and $\Omega_i = \varepsilon_{ijk} \frac{1}{8\pi} \vec n (\partial_j \vec n \times \partial_k \vec n)$.} positioned in the core on the intermediate planes $z =  m \lambda_h/2$ with $m \in \mathbb{Z}$. This screw dislocation sd$^{{\rm Bp}}_{1}$ with a chain of Bloch points at its core is the most stable configuration for $\nu = 1$ close to zero field but it is energetically more costly than the dislocations sd$^{\pm}_{-1}$ with $\nu = -1$. It can be maintained for a large field range but it becomes unstable in the simulations at point $C$ in Fig.~\ref{fig:energies} where oppositely charged Bloch points annihilate before reaching the critical field $H_{c2}$.  
 
There exist a third screw dislocation with strength $\nu = 1$ that again possesses a smooth core texture without singularities. Its core magnetization is either fully aligned or anti-aligned with the applied field. Both configurations, sd$^{{\rm Sk}+}_{1}$ and sd$^{{\rm Sk}-}_{1}$, respectively, are degenerate at zero field, but the anti-aligned core is energetically favoured at finite $H$, see Fig.~\ref{fig:energies}. In the simulations, the configuration with the aligned core can only be stabilized  for small field values up to point $D$. In order to elucidate the core structure, we focus in Fig.~\ref{Vd}(c) on sd$^{{\rm Sk}-}_{1}$ at $H=0$ and consider again the planes $z = \frac12(m \pm \frac{1}{2}) \lambda_h$ with a divergenceless configuration of the magnetization in the far field. As the core is approached within these planes, the magnetization smoothly rotates in a Bloch-like fashion that is favoured by the DMI. In half of these planes it is sufficient to rotate the magnetization by $\pi/2$ but in the complementary planes a rotation by $3\pi/2$ is required in order to reach a uniformly magnetized core. 
The topological charge within each plane is given by $N_{\rm top} = -\frac{1}{2} (1+H/H_{c2})$. As the field increases, the in-plane texture transforms from a meron with $N_{\rm top} = -\frac{1}{2}$ to a skyrmion with $N_{\rm top} = -1$. In fact, as the critical field $H_{c2}$ is approached this screw dislocation smoothly converts into a skyrmion configuration of the field-polarized state. The screw dislocation sd$^{{\rm Sk}}_{1}$ close to $H_{c2}$ can thus be viewed as a bound state of a skyrmion string with a vortex of the XY-order parameter $\psi$.

{\it Summary --} Screw dislocations belong to a diversity of topological textures in chiral magnets. They are line excitations of the conical helix phase and possess a universal far field that is characterized by their strength $\nu \in \mathbb{Z}$. Their micromagnetic core structure can be multifarious. Focusing on $\nu = \pm 1$ we showed that it can be either smooth or singular comprising a chain of Bloch points. The core of screw dislocations with larger strength $|\nu|$ can be even richer that will be shown elsewhere. Moreover, screw dislocations carry a finite non-integer skyrmion charge $N_{\rm top}$ and can thus be efficiently manipulated by spin currents \cite{Back_2020}.

M.G. is supported by the Deutsche Forschungsgemeinschaft (DFG, German Research Foundation) in the framework of SFB 1143 (project A07; project-id 247310070), project-id 270344603 and SPP 2137 (project-id 324327023).

\bibliography{screw_disclination}

\begin{thebibliography}{67}%
\makeatletter
\providecommand \@ifxundefined [1]{%
 \@ifx{#1\undefined}
}%
\providecommand \@ifnum [1]{%
 \ifnum #1\expandafter \@firstoftwo
 \else \expandafter \@secondoftwo
 \fi
}%
\providecommand \@ifx [1]{%
 \ifx #1\expandafter \@firstoftwo
 \else \expandafter \@secondoftwo
 \fi
}%
\providecommand \natexlab [1]{#1}%
\providecommand \enquote  [1]{``#1''}%
\providecommand \bibnamefont  [1]{#1}%
\providecommand \bibfnamefont [1]{#1}%
\providecommand \citenamefont [1]{#1}%
\providecommand \href@noop [0]{\@secondoftwo}%
\providecommand \href [0]{\begingroup \@sanitize@url \@href}%
\providecommand \@href[1]{\@@startlink{#1}\@@href}%
\providecommand \@@href[1]{\endgroup#1\@@endlink}%
\providecommand \@sanitize@url [0]{\catcode `\\12\catcode `\$12\catcode
  `\&12\catcode `\#12\catcode `\^12\catcode `\_12\catcode `\%12\relax}%
\providecommand \@@startlink[1]{}%
\providecommand \@@endlink[0]{}%
\providecommand \url  [0]{\begingroup\@sanitize@url \@url }%
\providecommand \@url [1]{\endgroup\@href {#1}{\urlprefix }}%
\providecommand \urlprefix  [0]{URL }%
\providecommand \Eprint [0]{\href }%
\providecommand \doibase [0]{https://doi.org/}%
\providecommand \selectlanguage [0]{\@gobble}%
\providecommand \bibinfo  [0]{\@secondoftwo}%
\providecommand \bibfield  [0]{\@secondoftwo}%
\providecommand \translation [1]{[#1]}%
\providecommand \BibitemOpen [0]{}%
\providecommand \bibitemStop [0]{}%
\providecommand \bibitemNoStop [0]{.\EOS\space}%
\providecommand \EOS [0]{\spacefactor3000\relax}%
\providecommand \BibitemShut  [1]{\csname bibitem#1\endcsname}%
\let\auto@bib@innerbib\@empty
\bibitem [{\citenamefont {Bak}\ and\ \citenamefont
  {Jensen}(1980)}]{BakJensen1980}%
  \BibitemOpen
  \bibfield  {author} {\bibinfo {author} {\bibfnamefont {P.}~\bibnamefont
  {Bak}}\ and\ \bibinfo {author} {\bibfnamefont {M.~H.}\ \bibnamefont
  {Jensen}},\ }\href {https://doi.org/10.1088/0022-3719/13/31/002} {\bibfield
  {journal} {\bibinfo  {journal} {Journal of Physics C: Solid State Physics}\
  }\textbf {\bibinfo {volume} {13}},\ \bibinfo {pages} {L881} (\bibinfo {year}
  {1980})}\BibitemShut {NoStop}%
\bibitem [{\citenamefont {Nakanishi}\ \emph {et~al.}(1980)\citenamefont
  {Nakanishi}, \citenamefont {Yanase}, \citenamefont {Hasegawa},\ and\
  \citenamefont {Kataoka}}]{Nakanishi1980}%
  \BibitemOpen
  \bibfield  {author} {\bibinfo {author} {\bibfnamefont {O.}~\bibnamefont
  {Nakanishi}}, \bibinfo {author} {\bibfnamefont {A.}~\bibnamefont {Yanase}},
  \bibinfo {author} {\bibfnamefont {A.}~\bibnamefont {Hasegawa}},\ and\
  \bibinfo {author} {\bibfnamefont {M.}~\bibnamefont {Kataoka}},\ }\href
  {https://doi.org/10.1016/0038-1098(80)91004-2} {\bibfield  {journal}
  {\bibinfo  {journal} {Solid State Communications}\ }\textbf {\bibinfo
  {volume} {35}},\ \bibinfo {pages} {995} (\bibinfo {year} {1980})}\BibitemShut
  {NoStop}%
\bibitem [{\citenamefont {Ishikawa}\ \emph {et~al.}(1976)\citenamefont
  {Ishikawa}, \citenamefont {Tajima}, \citenamefont {Bloch},\ and\
  \citenamefont {Roth}}]{Ishikawa1976}%
  \BibitemOpen
  \bibfield  {author} {\bibinfo {author} {\bibfnamefont {Y.}~\bibnamefont
  {Ishikawa}}, \bibinfo {author} {\bibfnamefont {K.}~\bibnamefont {Tajima}},
  \bibinfo {author} {\bibfnamefont {D.}~\bibnamefont {Bloch}},\ and\ \bibinfo
  {author} {\bibfnamefont {M.}~\bibnamefont {Roth}},\ }\href
  {https://doi.org/10.1016/0038-1098(76)90057-0} {\bibfield  {journal}
  {\bibinfo  {journal} {Solid State Communications}\ }\textbf {\bibinfo
  {volume} {19}},\ \bibinfo {pages} {525} (\bibinfo {year} {1976})}\BibitemShut
  {NoStop}%
\bibitem [{\citenamefont {Lebech}\ \emph {et~al.}(1989)\citenamefont {Lebech},
  \citenamefont {Bernhard},\ and\ \citenamefont {Freltoft}}]{Lebech1989}%
  \BibitemOpen
  \bibfield  {author} {\bibinfo {author} {\bibfnamefont {B.}~\bibnamefont
  {Lebech}}, \bibinfo {author} {\bibfnamefont {J.}~\bibnamefont {Bernhard}},\
  and\ \bibinfo {author} {\bibfnamefont {T.}~\bibnamefont {Freltoft}},\ }\href
  {https://doi.org/10.1088/0953-8984/1/35/010} {\bibfield  {journal} {\bibinfo
  {journal} {Journal of Physics: Condensed Matter}\ }\textbf {\bibinfo {volume}
  {1}},\ \bibinfo {pages} {6105} (\bibinfo {year} {1989})}\BibitemShut
  {NoStop}%
\bibitem [{\citenamefont {Uchida}\ \emph {et~al.}(2006)\citenamefont {Uchida},
  \citenamefont {Onose}, \citenamefont {Matsui},\ and\ \citenamefont
  {Tokura}}]{Uchida2006}%
  \BibitemOpen
  \bibfield  {author} {\bibinfo {author} {\bibfnamefont {M.}~\bibnamefont
  {Uchida}}, \bibinfo {author} {\bibfnamefont {Y.}~\bibnamefont {Onose}},
  \bibinfo {author} {\bibfnamefont {Y.}~\bibnamefont {Matsui}},\ and\ \bibinfo
  {author} {\bibfnamefont {Y.}~\bibnamefont {Tokura}},\ }\href
  {https://doi.org/10.1126/science.1120639} {\bibfield  {journal} {\bibinfo
  {journal} {Science}\ }\textbf {\bibinfo {volume} {311}},\ \bibinfo {pages}
  {359} (\bibinfo {year} {2006})}\BibitemShut {NoStop}%
\bibitem [{\citenamefont {Adams}\ \emph {et~al.}(2012)\citenamefont {Adams},
  \citenamefont {Chacon}, \citenamefont {Wagner}, \citenamefont {Bauer},
  \citenamefont {Brandl}, \citenamefont {Pedersen}, \citenamefont {Berger},
  \citenamefont {Lemmens},\ and\ \citenamefont {Pfleiderer}}]{Adams2012}%
  \BibitemOpen
  \bibfield  {author} {\bibinfo {author} {\bibfnamefont {T.}~\bibnamefont
  {Adams}}, \bibinfo {author} {\bibfnamefont {A.}~\bibnamefont {Chacon}},
  \bibinfo {author} {\bibfnamefont {M.}~\bibnamefont {Wagner}}, \bibinfo
  {author} {\bibfnamefont {A.}~\bibnamefont {Bauer}}, \bibinfo {author}
  {\bibfnamefont {G.}~\bibnamefont {Brandl}}, \bibinfo {author} {\bibfnamefont
  {B.}~\bibnamefont {Pedersen}}, \bibinfo {author} {\bibfnamefont
  {H.}~\bibnamefont {Berger}}, \bibinfo {author} {\bibfnamefont
  {P.}~\bibnamefont {Lemmens}},\ and\ \bibinfo {author} {\bibfnamefont
  {C.}~\bibnamefont {Pfleiderer}},\ }\href
  {https://doi.org/10.1103/PhysRevLett.108.237204} {\bibfield  {journal}
  {\bibinfo  {journal} {Physical Review Letters}\ }\textbf {\bibinfo {volume}
  {108}},\ \bibinfo {pages} {237204} (\bibinfo {year} {2012})}\BibitemShut
  {NoStop}%
\bibitem [{\citenamefont {Seki}\ \emph {et~al.}(2012)\citenamefont {Seki},
  \citenamefont {Yu}, \citenamefont {Ishiwata},\ and\ \citenamefont
  {Tokura}}]{Seki2012}%
  \BibitemOpen
  \bibfield  {author} {\bibinfo {author} {\bibfnamefont {S.}~\bibnamefont
  {Seki}}, \bibinfo {author} {\bibfnamefont {X.~Z.}\ \bibnamefont {Yu}},
  \bibinfo {author} {\bibfnamefont {S.}~\bibnamefont {Ishiwata}},\ and\
  \bibinfo {author} {\bibfnamefont {Y.}~\bibnamefont {Tokura}},\ }\href
  {https://doi.org/10.1126/science.1214143} {\bibfield  {journal} {\bibinfo
  {journal} {Science}\ }\textbf {\bibinfo {volume} {336}},\ \bibinfo {pages}
  {198} (\bibinfo {year} {2012})}\BibitemShut {NoStop}%
\bibitem [{\citenamefont {Chaikin}\ and\ \citenamefont
  {Lubensky}(2000)}]{Chaikin00}%
  \BibitemOpen
  \bibfield  {author} {\bibinfo {author} {\bibfnamefont {P.~M.}\ \bibnamefont
  {Chaikin}}\ and\ \bibinfo {author} {\bibfnamefont {T.~C.}\ \bibnamefont
  {Lubensky}},\ }\href@noop {} {\emph {\bibinfo {title} {Principles of
  Condensed Matter Physics}}},\ \bibinfo {edition} {1st}\ ed.\ (\bibinfo
  {publisher} {Cambridge University Press},\ \bibinfo {year}
  {2000})\BibitemShut {NoStop}%
\bibitem [{\citenamefont {de~Gennes}\ and\ \citenamefont
  {Prost}(1995)}]{P.G.deGennes95}%
  \BibitemOpen
  \bibfield  {author} {\bibinfo {author} {\bibfnamefont {P.~G.}\ \bibnamefont
  {de~Gennes}}\ and\ \bibinfo {author} {\bibfnamefont {J.}~\bibnamefont
  {Prost}},\ }\href
  {https://www.ebook.de/de/product/3237754/p_g_de_gennes_j_prost_the_physics_of_liquid_crystals.html}
  {\emph {\bibinfo {title} {The Physics of Liquid Crystals}}}\ (\bibinfo
  {publisher} {Oxford University Press},\ \bibinfo {year} {1995})\BibitemShut
  {NoStop}%
\bibitem [{\citenamefont {Kleman}\ and\ \citenamefont
  {Lavrentovich}(2004)}]{Kleman04}%
  \BibitemOpen
  \bibinfo {editor} {\bibfnamefont {M.}~\bibnamefont {Kleman}}\ and\ \bibinfo
  {editor} {\bibfnamefont {O.~D.}\ \bibnamefont {Lavrentovich}},\ eds.,\ \href
  {https://doi.org/10.1007/b97416} {\emph {\bibinfo {title} {Soft Matter
  Physics: An Introduction}}}\ (\bibinfo  {publisher} {Springer New York},\
  \bibinfo {year} {2004})\BibitemShut {NoStop}%
\bibitem [{\citenamefont {Janoschek}\ \emph {et~al.}(2013)\citenamefont
  {Janoschek}, \citenamefont {Garst}, \citenamefont {Bauer}, \citenamefont
  {Krautscheid}, \citenamefont {Georgii}, \citenamefont {B{\"o}ni},\ and\
  \citenamefont {Pfleiderer}}]{Janoschek2013}%
  \BibitemOpen
  \bibfield  {author} {\bibinfo {author} {\bibfnamefont {M.}~\bibnamefont
  {Janoschek}}, \bibinfo {author} {\bibfnamefont {M.}~\bibnamefont {Garst}},
  \bibinfo {author} {\bibfnamefont {A.}~\bibnamefont {Bauer}}, \bibinfo
  {author} {\bibfnamefont {P.}~\bibnamefont {Krautscheid}}, \bibinfo {author}
  {\bibfnamefont {R.}~\bibnamefont {Georgii}}, \bibinfo {author} {\bibfnamefont
  {P.}~\bibnamefont {B{\"o}ni}},\ and\ \bibinfo {author} {\bibfnamefont
  {C.}~\bibnamefont {Pfleiderer}},\ }\href
  {https://doi.org/10.1103/PhysRevB.87.134407} {\bibfield  {journal} {\bibinfo
  {journal} {Physical Review B}\ }\textbf {\bibinfo {volume} {87}},\ \bibinfo
  {pages} {134407} (\bibinfo {year} {2013})}\BibitemShut {NoStop}%
\bibitem [{\citenamefont {Buhrandt}\ and\ \citenamefont
  {Fritz}(2013)}]{Buhrandt2013}%
  \BibitemOpen
  \bibfield  {author} {\bibinfo {author} {\bibfnamefont {S.}~\bibnamefont
  {Buhrandt}}\ and\ \bibinfo {author} {\bibfnamefont {L.}~\bibnamefont
  {Fritz}},\ }\href {https://doi.org/10.1103/PhysRevB.88.195137} {\bibfield
  {journal} {\bibinfo  {journal} {Physical Review B}\ }\textbf {\bibinfo
  {volume} {88}},\ \bibinfo {pages} {195137} (\bibinfo {year}
  {2013})}\BibitemShut {NoStop}%
\bibitem [{\citenamefont {\ifmmode \check{Z}\else
  \v{Z}\fi{}ivkovi\ifmmode~\acute{c}\else \'{c}\fi{}}\ \emph
  {et~al.}(2014)\citenamefont {\ifmmode \check{Z}\else
  \v{Z}\fi{}ivkovi\ifmmode~\acute{c}\else \'{c}\fi{}}, \citenamefont {White},
  \citenamefont {R\o{}nnow}, \citenamefont {Pr\ifmmode~\check{s}\else
  \v{s}\fi{}a},\ and\ \citenamefont {Berger}}]{Zivkovic2014}%
  \BibitemOpen
  \bibfield  {author} {\bibinfo {author} {\bibfnamefont {I.}~\bibnamefont
  {\ifmmode \check{Z}\else \v{Z}\fi{}ivkovi\ifmmode~\acute{c}\else
  \'{c}\fi{}}}, \bibinfo {author} {\bibfnamefont {J.~S.}\ \bibnamefont
  {White}}, \bibinfo {author} {\bibfnamefont {H.~M.}\ \bibnamefont
  {R\o{}nnow}}, \bibinfo {author} {\bibfnamefont {K.}~\bibnamefont
  {Pr\ifmmode~\check{s}\else \v{s}\fi{}a}},\ and\ \bibinfo {author}
  {\bibfnamefont {H.}~\bibnamefont {Berger}},\ }\href
  {https://doi.org/10.1103/PhysRevB.89.060401} {\bibfield  {journal} {\bibinfo
  {journal} {Phys. Rev. B}\ }\textbf {\bibinfo {volume} {89}},\ \bibinfo
  {pages} {060401} (\bibinfo {year} {2014})}\BibitemShut {NoStop}%
\bibitem [{\citenamefont {Brazovskii}\ \emph {et~al.}(1987)\citenamefont
  {Brazovskii}, \citenamefont {Dzyaloshinskii},\ and\ \citenamefont
  {Muratov}}]{Brazovskii1987}%
  \BibitemOpen
  \bibfield  {author} {\bibinfo {author} {\bibfnamefont {S.~A.}\ \bibnamefont
  {Brazovskii}}, \bibinfo {author} {\bibfnamefont {I.~E.}\ \bibnamefont
  {Dzyaloshinskii}},\ and\ \bibinfo {author} {\bibfnamefont {A.~R.}\
  \bibnamefont {Muratov}},\ }\href@noop {} {\bibfield  {journal} {\bibinfo
  {journal} {Sov. Phys. JETP}\ ,\ \bibinfo {pages} {625}} (\bibinfo {year}
  {1987})}\BibitemShut {NoStop}%
\bibitem [{\citenamefont {Kindervater}\ \emph {et~al.}(2019)\citenamefont
  {Kindervater}, \citenamefont {Stasinopoulos}, \citenamefont {Bauer},
  \citenamefont {Haslbeck}, \citenamefont {Rucker}, \citenamefont {Chacon},
  \citenamefont {M\"uhlbauer}, \citenamefont {Franz}, \citenamefont {Garst},
  \citenamefont {Grundler},\ and\ \citenamefont
  {Pfleiderer}}]{Kindervater2019}%
  \BibitemOpen
  \bibfield  {author} {\bibinfo {author} {\bibfnamefont {J.}~\bibnamefont
  {Kindervater}}, \bibinfo {author} {\bibfnamefont {I.}~\bibnamefont
  {Stasinopoulos}}, \bibinfo {author} {\bibfnamefont {A.}~\bibnamefont
  {Bauer}}, \bibinfo {author} {\bibfnamefont {F.~X.}\ \bibnamefont {Haslbeck}},
  \bibinfo {author} {\bibfnamefont {F.}~\bibnamefont {Rucker}}, \bibinfo
  {author} {\bibfnamefont {A.}~\bibnamefont {Chacon}}, \bibinfo {author}
  {\bibfnamefont {S.}~\bibnamefont {M\"uhlbauer}}, \bibinfo {author}
  {\bibfnamefont {C.}~\bibnamefont {Franz}}, \bibinfo {author} {\bibfnamefont
  {M.}~\bibnamefont {Garst}}, \bibinfo {author} {\bibfnamefont
  {D.}~\bibnamefont {Grundler}},\ and\ \bibinfo {author} {\bibfnamefont
  {C.}~\bibnamefont {Pfleiderer}},\ }\href
  {https://doi.org/10.1103/PhysRevX.9.041059} {\bibfield  {journal} {\bibinfo
  {journal} {Physical Review X}\ }\textbf {\bibinfo {volume} {9}},\ \bibinfo
  {pages} {041059} (\bibinfo {year} {2019})}\BibitemShut {NoStop}%
\bibitem [{\citenamefont {Pfleiderer}\ \emph {et~al.}(2001)\citenamefont
  {Pfleiderer}, \citenamefont {Julian},\ and\ \citenamefont
  {Lonzarich}}]{Pfleiderer2001}%
  \BibitemOpen
  \bibfield  {author} {\bibinfo {author} {\bibfnamefont {C.}~\bibnamefont
  {Pfleiderer}}, \bibinfo {author} {\bibfnamefont {S.~R.}\ \bibnamefont
  {Julian}},\ and\ \bibinfo {author} {\bibfnamefont {G.~G.}\ \bibnamefont
  {Lonzarich}},\ }\href {https://doi.org/10.1038/35106527} {\bibfield
  {journal} {\bibinfo  {journal} {Nature}\ }\textbf {\bibinfo {volume} {414}},\
  \bibinfo {pages} {427} (\bibinfo {year} {2001})}\BibitemShut {NoStop}%
\bibitem [{\citenamefont {Pfleiderer}\ \emph {et~al.}(2007)\citenamefont
  {Pfleiderer}, \citenamefont {Boni}, \citenamefont {Keller}, \citenamefont
  {Rossler},\ and\ \citenamefont {Rosch}}]{Pfleiderer2007}%
  \BibitemOpen
  \bibfield  {author} {\bibinfo {author} {\bibfnamefont {C.}~\bibnamefont
  {Pfleiderer}}, \bibinfo {author} {\bibfnamefont {P.}~\bibnamefont {Boni}},
  \bibinfo {author} {\bibfnamefont {T.}~\bibnamefont {Keller}}, \bibinfo
  {author} {\bibfnamefont {U.~K.}\ \bibnamefont {Rossler}},\ and\ \bibinfo
  {author} {\bibfnamefont {A.}~\bibnamefont {Rosch}},\ }\href
  {https://doi.org/10.1126/science.1142644} {\bibfield  {journal} {\bibinfo
  {journal} {Science}\ }\textbf {\bibinfo {volume} {316}},\ \bibinfo {pages}
  {1871} (\bibinfo {year} {2007})}\BibitemShut {NoStop}%
\bibitem [{\citenamefont {Pedrazzini}\ \emph {et~al.}(2007)\citenamefont
  {Pedrazzini}, \citenamefont {Wilhelm}, \citenamefont {Jaccard}, \citenamefont
  {Jarlborg}, \citenamefont {Schmidt}, \citenamefont {Hanfland}, \citenamefont
  {Akselrud}, \citenamefont {Yuan}, \citenamefont {Schwarz}, \citenamefont
  {Grin},\ and\ \citenamefont {Steglich}}]{Pedrazzini2007}%
  \BibitemOpen
  \bibfield  {author} {\bibinfo {author} {\bibfnamefont {P.}~\bibnamefont
  {Pedrazzini}}, \bibinfo {author} {\bibfnamefont {H.}~\bibnamefont {Wilhelm}},
  \bibinfo {author} {\bibfnamefont {D.}~\bibnamefont {Jaccard}}, \bibinfo
  {author} {\bibfnamefont {T.}~\bibnamefont {Jarlborg}}, \bibinfo {author}
  {\bibfnamefont {M.}~\bibnamefont {Schmidt}}, \bibinfo {author} {\bibfnamefont
  {M.}~\bibnamefont {Hanfland}}, \bibinfo {author} {\bibfnamefont
  {L.}~\bibnamefont {Akselrud}}, \bibinfo {author} {\bibfnamefont {H.~Q.}\
  \bibnamefont {Yuan}}, \bibinfo {author} {\bibfnamefont {U.}~\bibnamefont
  {Schwarz}}, \bibinfo {author} {\bibfnamefont {Y.}~\bibnamefont {Grin}},\ and\
  \bibinfo {author} {\bibfnamefont {F.}~\bibnamefont {Steglich}},\ }\href
  {https://doi.org/10.1103/PhysRevLett.98.047204} {\bibfield  {journal}
  {\bibinfo  {journal} {Physical Review Letters}\ }\textbf {\bibinfo {volume}
  {98}},\ \bibinfo {pages} {047204} (\bibinfo {year} {2007})}\BibitemShut
  {NoStop}%
\bibitem [{\citenamefont {Ritz}\ \emph {et~al.}(2013)\citenamefont {Ritz},
  \citenamefont {Halder}, \citenamefont {Wagner}, \citenamefont {Franz},
  \citenamefont {Bauer},\ and\ \citenamefont {Pfleiderer}}]{Ritz2013}%
  \BibitemOpen
  \bibfield  {author} {\bibinfo {author} {\bibfnamefont {R.}~\bibnamefont
  {Ritz}}, \bibinfo {author} {\bibfnamefont {M.}~\bibnamefont {Halder}},
  \bibinfo {author} {\bibfnamefont {M.}~\bibnamefont {Wagner}}, \bibinfo
  {author} {\bibfnamefont {C.}~\bibnamefont {Franz}}, \bibinfo {author}
  {\bibfnamefont {A.}~\bibnamefont {Bauer}},\ and\ \bibinfo {author}
  {\bibfnamefont {C.}~\bibnamefont {Pfleiderer}},\ }\href
  {https://doi.org/10.1038/nature12023} {\bibfield  {journal} {\bibinfo
  {journal} {Nature}\ }\textbf {\bibinfo {volume} {497}},\ \bibinfo {pages}
  {231} (\bibinfo {year} {2013})}\BibitemShut {NoStop}%
\bibitem [{\citenamefont {Schmalian}\ and\ \citenamefont
  {Turlakov}(2004)}]{Schmalian2004}%
  \BibitemOpen
  \bibfield  {author} {\bibinfo {author} {\bibfnamefont {J.}~\bibnamefont
  {Schmalian}}\ and\ \bibinfo {author} {\bibfnamefont {M.}~\bibnamefont
  {Turlakov}},\ }\href {https://doi.org/10.1103/PhysRevLett.93.036405}
  {\bibfield  {journal} {\bibinfo  {journal} {Physical Review Letters}\
  }\textbf {\bibinfo {volume} {93}},\ \bibinfo {pages} {036405} (\bibinfo
  {year} {2004})}\BibitemShut {NoStop}%
\bibitem [{\citenamefont {Toner}(1982)}]{Toner1982}%
  \BibitemOpen
  \bibfield  {author} {\bibinfo {author} {\bibfnamefont {J.}~\bibnamefont
  {Toner}},\ }\href {https://doi.org/10.1103/PhysRevB.26.462} {\bibfield
  {journal} {\bibinfo  {journal} {Physical Review B}\ }\textbf {\bibinfo
  {volume} {26}},\ \bibinfo {pages} {462} (\bibinfo {year} {1982})}\BibitemShut
  {NoStop}%
\bibitem [{\citenamefont {Grinstein}\ \emph {et~al.}(1986)\citenamefont
  {Grinstein}, \citenamefont {Lubensky},\ and\ \citenamefont
  {Toner}}]{Grinstein1986}%
  \BibitemOpen
  \bibfield  {author} {\bibinfo {author} {\bibfnamefont {G.}~\bibnamefont
  {Grinstein}}, \bibinfo {author} {\bibfnamefont {T.~C.}\ \bibnamefont
  {Lubensky}},\ and\ \bibinfo {author} {\bibfnamefont {J.}~\bibnamefont
  {Toner}},\ }\href {https://doi.org/10.1103/PhysRevB.33.3306} {\bibfield
  {journal} {\bibinfo  {journal} {Physical Review B}\ }\textbf {\bibinfo
  {volume} {33}},\ \bibinfo {pages} {3306} (\bibinfo {year}
  {1986})}\BibitemShut {NoStop}%
\bibitem [{\citenamefont {Zhai}\ and\ \citenamefont
  {Radzihovsky}(2021)}]{Zhai2021}%
  \BibitemOpen
  \bibfield  {author} {\bibinfo {author} {\bibfnamefont {Z.}~\bibnamefont
  {Zhai}}\ and\ \bibinfo {author} {\bibfnamefont {L.}~\bibnamefont
  {Radzihovsky}},\ }\bibfield  {journal} {\bibinfo  {journal} {Annals of
  Physics}\ }\href {https://doi.org/10.1016/j.aop.2021.168509}
  {10.1016/j.aop.2021.168509} (\bibinfo {year} {2021})\BibitemShut {NoStop}%
\bibitem [{\citenamefont {Kl{\'{e}}man}(1970)}]{Kleman70}%
  \BibitemOpen
  \bibfield  {author} {\bibinfo {author} {\bibfnamefont {M.}~\bibnamefont
  {Kl{\'{e}}man}},\ }\href {https://doi.org/10.1080/14786437008220943}
  {\bibfield  {journal} {\bibinfo  {journal} {Philosophical Magazine}\ }\textbf
  {\bibinfo {volume} {22}},\ \bibinfo {pages} {739} (\bibinfo {year}
  {1970})}\BibitemShut {NoStop}%
\bibitem [{\citenamefont {Li}\ \emph {et~al.}(2012)\citenamefont {Li},
  \citenamefont {Nattermann},\ and\ \citenamefont {Pokrovsky}}]{Li2012}%
  \BibitemOpen
  \bibfield  {author} {\bibinfo {author} {\bibfnamefont {F.}~\bibnamefont
  {Li}}, \bibinfo {author} {\bibfnamefont {T.}~\bibnamefont {Nattermann}},\
  and\ \bibinfo {author} {\bibfnamefont {V.~L.}\ \bibnamefont {Pokrovsky}},\
  }\href {https://doi.org/10.1103/PhysRevLett.108.107203} {\bibfield  {journal}
  {\bibinfo  {journal} {Phys. Rev. Lett.}\ }\textbf {\bibinfo {volume} {108}},\
  \bibinfo {pages} {107203} (\bibinfo {year} {2012})}\BibitemShut {NoStop}%
\bibitem [{\citenamefont {Nattermann}\ and\ \citenamefont
  {Pokrovsky}(2018)}]{Nattermann2018}%
  \BibitemOpen
  \bibfield  {author} {\bibinfo {author} {\bibfnamefont {T.}~\bibnamefont
  {Nattermann}}\ and\ \bibinfo {author} {\bibfnamefont {V.~L.}\ \bibnamefont
  {Pokrovsky}},\ }\href {https://doi.org/10.1134/S106377611811016X} {\bibfield
  {journal} {\bibinfo  {journal} {Journal of Experimental and Theoretical
  Physics}\ }\textbf {\bibinfo {volume} {127}},\ \bibinfo {pages} {922}
  (\bibinfo {year} {2018})}\BibitemShut {NoStop}%
\bibitem [{\citenamefont {Schoenherr}\ \emph {et~al.}(2018)\citenamefont
  {Schoenherr}, \citenamefont {Müller}, \citenamefont {Köhler}, \citenamefont
  {Rosch}, \citenamefont {Kanazawa}, \citenamefont {Tokura}, \citenamefont
  {Garst},\ and\ \citenamefont {Meier}}]{Schoenherr18}%
  \BibitemOpen
  \bibfield  {author} {\bibinfo {author} {\bibfnamefont {P.}~\bibnamefont
  {Schoenherr}}, \bibinfo {author} {\bibfnamefont {J.}~\bibnamefont {Müller}},
  \bibinfo {author} {\bibfnamefont {L.}~\bibnamefont {Köhler}}, \bibinfo
  {author} {\bibfnamefont {A.}~\bibnamefont {Rosch}}, \bibinfo {author}
  {\bibfnamefont {N.}~\bibnamefont {Kanazawa}}, \bibinfo {author}
  {\bibfnamefont {Y.}~\bibnamefont {Tokura}}, \bibinfo {author} {\bibfnamefont
  {M.}~\bibnamefont {Garst}},\ and\ \bibinfo {author} {\bibfnamefont
  {D.}~\bibnamefont {Meier}},\ }\href
  {https://doi.org/10.1038/s41567-018-0056-5} {\bibfield  {journal} {\bibinfo
  {journal} {Nature Physics}\ }\textbf {\bibinfo {volume} {14}},\ \bibinfo
  {pages} {465} (\bibinfo {year} {2018})}\BibitemShut {NoStop}%
\bibitem [{\citenamefont {Bouligand}(1983)}]{Bouligand1983}%
  \BibitemOpen
  \bibfield  {author} {\bibinfo {author} {\bibfnamefont {Y.}~\bibnamefont
  {Bouligand}},\ }\bibinfo {title} {Dislocations in solids}\ (\bibinfo
  {publisher} {North-Holland Publishing Company},\ \bibinfo {address} {New
  York},\ \bibinfo {year} {1983})\ Chap.~\bibinfo {chapter} {23}\BibitemShut
  {NoStop}%
\bibitem [{\citenamefont {Martin}\ \emph {et~al.}(2017)\citenamefont {Martin},
  \citenamefont {Deutsch}, \citenamefont {Chaboussant}, \citenamefont {Damay},
  \citenamefont {Bonville}, \citenamefont {Fomicheva}, \citenamefont
  {Tsvyashchenko}, \citenamefont {R\"ossler},\ and\ \citenamefont
  {Mirebeau}}]{Martin2017}%
  \BibitemOpen
  \bibfield  {author} {\bibinfo {author} {\bibfnamefont {N.}~\bibnamefont
  {Martin}}, \bibinfo {author} {\bibfnamefont {M.}~\bibnamefont {Deutsch}},
  \bibinfo {author} {\bibfnamefont {G.}~\bibnamefont {Chaboussant}}, \bibinfo
  {author} {\bibfnamefont {F.}~\bibnamefont {Damay}}, \bibinfo {author}
  {\bibfnamefont {P.}~\bibnamefont {Bonville}}, \bibinfo {author}
  {\bibfnamefont {L.~N.}\ \bibnamefont {Fomicheva}}, \bibinfo {author}
  {\bibfnamefont {A.~V.}\ \bibnamefont {Tsvyashchenko}}, \bibinfo {author}
  {\bibfnamefont {U.~K.}\ \bibnamefont {R\"ossler}},\ and\ \bibinfo {author}
  {\bibfnamefont {I.}~\bibnamefont {Mirebeau}},\ }\href
  {https://doi.org/10.1103/PhysRevB.96.020413} {\bibfield  {journal} {\bibinfo
  {journal} {Phys. Rev. B}\ }\textbf {\bibinfo {volume} {96}},\ \bibinfo
  {pages} {020413} (\bibinfo {year} {2017})}\BibitemShut {NoStop}%
\bibitem [{\citenamefont {Bauer}\ \emph {et~al.}(2017)\citenamefont {Bauer},
  \citenamefont {Chacon}, \citenamefont {Wagner}, \citenamefont {Halder},
  \citenamefont {Georgii}, \citenamefont {Rosch}, \citenamefont {Pfleiderer},\
  and\ \citenamefont {Garst}}]{Bauer2017}%
  \BibitemOpen
  \bibfield  {author} {\bibinfo {author} {\bibfnamefont {A.}~\bibnamefont
  {Bauer}}, \bibinfo {author} {\bibfnamefont {A.}~\bibnamefont {Chacon}},
  \bibinfo {author} {\bibfnamefont {M.}~\bibnamefont {Wagner}}, \bibinfo
  {author} {\bibfnamefont {M.}~\bibnamefont {Halder}}, \bibinfo {author}
  {\bibfnamefont {R.}~\bibnamefont {Georgii}}, \bibinfo {author} {\bibfnamefont
  {A.}~\bibnamefont {Rosch}}, \bibinfo {author} {\bibfnamefont
  {C.}~\bibnamefont {Pfleiderer}},\ and\ \bibinfo {author} {\bibfnamefont
  {M.}~\bibnamefont {Garst}},\ }\href
  {https://doi.org/10.1103/PhysRevB.95.024429} {\bibfield  {journal} {\bibinfo
  {journal} {Physical Review B}\ }\textbf {\bibinfo {volume} {95}},\ \bibinfo
  {pages} {024429} (\bibinfo {year} {2017})}\BibitemShut {NoStop}%
\bibitem [{\citenamefont {Milde}\ \emph {et~al.}(2020)\citenamefont {Milde},
  \citenamefont {Köhler}, \citenamefont {Neuber}, \citenamefont {Ritzinger},
  \citenamefont {Garst}, \citenamefont {Bauer}, \citenamefont {Pfleiderer},
  \citenamefont {Berger},\ and\ \citenamefont {Eng}}]{Milde2020}%
  \BibitemOpen
  \bibfield  {author} {\bibinfo {author} {\bibfnamefont {P.}~\bibnamefont
  {Milde}}, \bibinfo {author} {\bibfnamefont {L.}~\bibnamefont {Köhler}},
  \bibinfo {author} {\bibfnamefont {E.}~\bibnamefont {Neuber}}, \bibinfo
  {author} {\bibfnamefont {P.}~\bibnamefont {Ritzinger}}, \bibinfo {author}
  {\bibfnamefont {M.}~\bibnamefont {Garst}}, \bibinfo {author} {\bibfnamefont
  {A.}~\bibnamefont {Bauer}}, \bibinfo {author} {\bibfnamefont
  {C.}~\bibnamefont {Pfleiderer}}, \bibinfo {author} {\bibfnamefont
  {H.}~\bibnamefont {Berger}},\ and\ \bibinfo {author} {\bibfnamefont {L.~M.}\
  \bibnamefont {Eng}},\ }\href {https://doi.org/10.1103/PhysRevB.102.024426}
  {\bibfield  {journal} {\bibinfo  {journal} {Physical Review B}\ }\textbf
  {\bibinfo {volume} {102}},\ \bibinfo {pages} {024426} (\bibinfo {year}
  {2020})}\BibitemShut {NoStop}%
\bibitem [{\citenamefont {Dussaux}\ \emph {et~al.}(2016)\citenamefont
  {Dussaux}, \citenamefont {Schoenherr}, \citenamefont {Koumpouras},
  \citenamefont {Chico}, \citenamefont {Chang}, \citenamefont {Lorenzelli},
  \citenamefont {Kanazawa}, \citenamefont {Tokura}, \citenamefont {Garst},
  \citenamefont {Bergman}, \citenamefont {Degen},\ and\ \citenamefont
  {Meier}}]{Dussaux16}%
  \BibitemOpen
  \bibfield  {author} {\bibinfo {author} {\bibfnamefont {A.}~\bibnamefont
  {Dussaux}}, \bibinfo {author} {\bibfnamefont {P.}~\bibnamefont {Schoenherr}},
  \bibinfo {author} {\bibfnamefont {K.}~\bibnamefont {Koumpouras}}, \bibinfo
  {author} {\bibfnamefont {J.}~\bibnamefont {Chico}}, \bibinfo {author}
  {\bibfnamefont {K.}~\bibnamefont {Chang}}, \bibinfo {author} {\bibfnamefont
  {L.}~\bibnamefont {Lorenzelli}}, \bibinfo {author} {\bibfnamefont
  {N.}~\bibnamefont {Kanazawa}}, \bibinfo {author} {\bibfnamefont
  {Y.}~\bibnamefont {Tokura}}, \bibinfo {author} {\bibfnamefont
  {M.}~\bibnamefont {Garst}}, \bibinfo {author} {\bibfnamefont
  {A.}~\bibnamefont {Bergman}}, \bibinfo {author} {\bibfnamefont {C.~L.}\
  \bibnamefont {Degen}},\ and\ \bibinfo {author} {\bibfnamefont
  {D.}~\bibnamefont {Meier}},\ }\href {https://doi.org/10.1038/ncomms12430}
  {\bibfield  {journal} {\bibinfo  {journal} {Nature Communications}\ }\textbf
  {\bibinfo {volume} {7}},\ \bibinfo {pages} {12430} (\bibinfo {year}
  {2016})}\BibitemShut {NoStop}%
\bibitem [{\citenamefont {Schoenherr}\ \emph {et~al.}(2021)\citenamefont
  {Schoenherr}, \citenamefont {Stepanova}, \citenamefont {Lysne}, \citenamefont
  {Kanazawa}, \citenamefont {Tokura}, \citenamefont {Bergman},\ and\
  \citenamefont {Meier}}]{schoenherr2021}%
  \BibitemOpen
  \bibfield  {author} {\bibinfo {author} {\bibfnamefont {P.}~\bibnamefont
  {Schoenherr}}, \bibinfo {author} {\bibfnamefont {M.}~\bibnamefont
  {Stepanova}}, \bibinfo {author} {\bibfnamefont {E.~N.}\ \bibnamefont
  {Lysne}}, \bibinfo {author} {\bibfnamefont {N.}~\bibnamefont {Kanazawa}},
  \bibinfo {author} {\bibfnamefont {Y.}~\bibnamefont {Tokura}}, \bibinfo
  {author} {\bibfnamefont {A.}~\bibnamefont {Bergman}},\ and\ \bibinfo {author}
  {\bibfnamefont {D.}~\bibnamefont {Meier}},\ }\href@noop {} {} (\bibinfo
  {year} {2021}),\ \Eprint {https://arxiv.org/abs/2105.00658} {arXiv:2105.00658
  [cond-mat.mtrl-sci]} \BibitemShut {NoStop}%
\bibitem [{\citenamefont {Bogdanov}\ and\ \citenamefont
  {Hubert}(1994)}]{Bogdanov1994}%
  \BibitemOpen
  \bibfield  {author} {\bibinfo {author} {\bibfnamefont {A.}~\bibnamefont
  {Bogdanov}}\ and\ \bibinfo {author} {\bibfnamefont {A.}~\bibnamefont
  {Hubert}},\ }\href {https://doi.org/10.1016/0304-8853(94)90046-9} {\bibfield
  {journal} {\bibinfo  {journal} {Journal of Magnetism and Magnetic Materials}\
  }\textbf {\bibinfo {volume} {138}},\ \bibinfo {pages} {255} (\bibinfo {year}
  {1994})}\BibitemShut {NoStop}%
\bibitem [{\citenamefont {M\"uhlbauer}\ \emph {et~al.}(2009)\citenamefont
  {M\"uhlbauer}, \citenamefont {Binz}, \citenamefont {Jonietz}, \citenamefont
  {Pfleiderer}, \citenamefont {Rosch}, \citenamefont {Neubauer}, \citenamefont
  {Georgii},\ and\ \citenamefont {B\"oni}}]{Muhlbauer2009}%
  \BibitemOpen
  \bibfield  {author} {\bibinfo {author} {\bibfnamefont {S.}~\bibnamefont
  {M\"uhlbauer}}, \bibinfo {author} {\bibfnamefont {B.}~\bibnamefont {Binz}},
  \bibinfo {author} {\bibfnamefont {F.}~\bibnamefont {Jonietz}}, \bibinfo
  {author} {\bibfnamefont {C.}~\bibnamefont {Pfleiderer}}, \bibinfo {author}
  {\bibfnamefont {A.}~\bibnamefont {Rosch}}, \bibinfo {author} {\bibfnamefont
  {A.}~\bibnamefont {Neubauer}}, \bibinfo {author} {\bibfnamefont
  {R.}~\bibnamefont {Georgii}},\ and\ \bibinfo {author} {\bibfnamefont
  {P.}~\bibnamefont {B\"oni}},\ }\href
  {https://doi.org/10.1126/science.1166767} {\bibfield  {journal} {\bibinfo
  {journal} {Science}\ }\textbf {\bibinfo {volume} {323}},\ \bibinfo {pages}
  {915} (\bibinfo {year} {2009})}\BibitemShut {NoStop}%
\bibitem [{\citenamefont {Yu}\ \emph {et~al.}(2010)\citenamefont {Yu},
  \citenamefont {Onose}, \citenamefont {Kanazawa}, \citenamefont {Park},
  \citenamefont {Han}, \citenamefont {Matsui}, \citenamefont {Nagaosa},\ and\
  \citenamefont {Tokura}}]{Yu2010}%
  \BibitemOpen
  \bibfield  {author} {\bibinfo {author} {\bibfnamefont {X.~Z.}\ \bibnamefont
  {Yu}}, \bibinfo {author} {\bibfnamefont {Y.}~\bibnamefont {Onose}}, \bibinfo
  {author} {\bibfnamefont {N.}~\bibnamefont {Kanazawa}}, \bibinfo {author}
  {\bibfnamefont {J.~H.}\ \bibnamefont {Park}}, \bibinfo {author}
  {\bibfnamefont {J.~H.}\ \bibnamefont {Han}}, \bibinfo {author} {\bibfnamefont
  {Y.}~\bibnamefont {Matsui}}, \bibinfo {author} {\bibfnamefont
  {N.}~\bibnamefont {Nagaosa}},\ and\ \bibinfo {author} {\bibfnamefont
  {Y.}~\bibnamefont {Tokura}},\ }\href {https://doi.org/10.1038/nature09124}
  {\bibfield  {journal} {\bibinfo  {journal} {Nature}\ }\textbf {\bibinfo
  {volume} {465}},\ \bibinfo {pages} {901} (\bibinfo {year}
  {2010})}\BibitemShut {NoStop}%
\bibitem [{\citenamefont {Back}\ \emph {et~al.}(2020)\citenamefont {Back},
  \citenamefont {Cros}, \citenamefont {Ebert}, \citenamefont {Everschor-Sitte},
  \citenamefont {Fert}, \citenamefont {Garst}, \citenamefont {Ma},
  \citenamefont {Mankovsky}, \citenamefont {Monchesky}, \citenamefont
  {Mostovoy}, \citenamefont {Nagaosa}, \citenamefont {Parkin}, \citenamefont
  {Pfleiderer}, \citenamefont {Reyren}, \citenamefont {Rosch}, \citenamefont
  {Taguchi}, \citenamefont {Tokura}, \citenamefont {von Bergmann},\ and\
  \citenamefont {Zang}}]{Back_2020}%
  \BibitemOpen
  \bibfield  {author} {\bibinfo {author} {\bibfnamefont {C.}~\bibnamefont
  {Back}}, \bibinfo {author} {\bibfnamefont {V.}~\bibnamefont {Cros}}, \bibinfo
  {author} {\bibfnamefont {H.}~\bibnamefont {Ebert}}, \bibinfo {author}
  {\bibfnamefont {K.}~\bibnamefont {Everschor-Sitte}}, \bibinfo {author}
  {\bibfnamefont {A.}~\bibnamefont {Fert}}, \bibinfo {author} {\bibfnamefont
  {M.}~\bibnamefont {Garst}}, \bibinfo {author} {\bibfnamefont
  {T.}~\bibnamefont {Ma}}, \bibinfo {author} {\bibfnamefont {S.}~\bibnamefont
  {Mankovsky}}, \bibinfo {author} {\bibfnamefont {T.~L.}\ \bibnamefont
  {Monchesky}}, \bibinfo {author} {\bibfnamefont {M.}~\bibnamefont {Mostovoy}},
  \bibinfo {author} {\bibfnamefont {N.}~\bibnamefont {Nagaosa}}, \bibinfo
  {author} {\bibfnamefont {S.~S.~P.}\ \bibnamefont {Parkin}}, \bibinfo {author}
  {\bibfnamefont {C.}~\bibnamefont {Pfleiderer}}, \bibinfo {author}
  {\bibfnamefont {N.}~\bibnamefont {Reyren}}, \bibinfo {author} {\bibfnamefont
  {A.}~\bibnamefont {Rosch}}, \bibinfo {author} {\bibfnamefont
  {Y.}~\bibnamefont {Taguchi}}, \bibinfo {author} {\bibfnamefont
  {Y.}~\bibnamefont {Tokura}}, \bibinfo {author} {\bibfnamefont
  {K.}~\bibnamefont {von Bergmann}},\ and\ \bibinfo {author} {\bibfnamefont
  {J.}~\bibnamefont {Zang}},\ }\href {https://doi.org/10.1088/1361-6463/ab8418}
  {\bibfield  {journal} {\bibinfo  {journal} {Journal of Physics D: Applied
  Physics}\ }\textbf {\bibinfo {volume} {53}},\ \bibinfo {pages} {363001}
  (\bibinfo {year} {2020})}\BibitemShut {NoStop}%
\bibitem [{\citenamefont {Leonov}\ \emph {et~al.}(2016)\citenamefont {Leonov},
  \citenamefont {Monchesky}, \citenamefont {Loudon},\ and\ \citenamefont
  {Bogdanov}}]{Leonov16a}%
  \BibitemOpen
  \bibfield  {author} {\bibinfo {author} {\bibfnamefont {A.~O.}\ \bibnamefont
  {Leonov}}, \bibinfo {author} {\bibfnamefont {T.~L.}\ \bibnamefont
  {Monchesky}}, \bibinfo {author} {\bibfnamefont {J.~C.}\ \bibnamefont
  {Loudon}},\ and\ \bibinfo {author} {\bibfnamefont {A.~N.}\ \bibnamefont
  {Bogdanov}},\ }\href {https://doi.org/10.1088/0953-8984/28/35/35lt01}
  {\bibfield  {journal} {\bibinfo  {journal} {Journal of Physics: Condensed
  Matter}\ }\textbf {\bibinfo {volume} {28}},\ \bibinfo {pages} {35LT01}
  (\bibinfo {year} {2016})}\BibitemShut {NoStop}%
\bibitem [{\citenamefont {Du}\ \emph {et~al.}(2018)\citenamefont {Du},
  \citenamefont {Zhao}, \citenamefont {Rybakov}, \citenamefont {Borisov},
  \citenamefont {Wang}, \citenamefont {Tang}, \citenamefont {Jin},
  \citenamefont {Wang}, \citenamefont {Wei}, \citenamefont {Kiselev},
  \citenamefont {Zhang}, \citenamefont {Che}, \citenamefont {Bl\"ugel},\ and\
  \citenamefont {Tian}}]{Du2018}%
  \BibitemOpen
  \bibfield  {author} {\bibinfo {author} {\bibfnamefont {H.}~\bibnamefont
  {Du}}, \bibinfo {author} {\bibfnamefont {X.}~\bibnamefont {Zhao}}, \bibinfo
  {author} {\bibfnamefont {F.~N.}\ \bibnamefont {Rybakov}}, \bibinfo {author}
  {\bibfnamefont {A.~B.}\ \bibnamefont {Borisov}}, \bibinfo {author}
  {\bibfnamefont {S.}~\bibnamefont {Wang}}, \bibinfo {author} {\bibfnamefont
  {J.}~\bibnamefont {Tang}}, \bibinfo {author} {\bibfnamefont {C.}~\bibnamefont
  {Jin}}, \bibinfo {author} {\bibfnamefont {C.}~\bibnamefont {Wang}}, \bibinfo
  {author} {\bibfnamefont {W.}~\bibnamefont {Wei}}, \bibinfo {author}
  {\bibfnamefont {N.~S.}\ \bibnamefont {Kiselev}}, \bibinfo {author}
  {\bibfnamefont {Y.}~\bibnamefont {Zhang}}, \bibinfo {author} {\bibfnamefont
  {R.}~\bibnamefont {Che}}, \bibinfo {author} {\bibfnamefont {S.}~\bibnamefont
  {Bl\"ugel}},\ and\ \bibinfo {author} {\bibfnamefont {M.}~\bibnamefont
  {Tian}},\ }\href {https://doi.org/10.1103/PhysRevLett.120.197203} {\bibfield
  {journal} {\bibinfo  {journal} {Phys. Rev. Lett.}\ }\textbf {\bibinfo
  {volume} {120}},\ \bibinfo {pages} {197203} (\bibinfo {year}
  {2018})}\BibitemShut {NoStop}%
\bibitem [{\citenamefont {Sohn}\ \emph {et~al.}(2019)\citenamefont {Sohn},
  \citenamefont {Vlasov}, \citenamefont {Uzdin}, \citenamefont {Leonov},\ and\
  \citenamefont {Smalyukh}}]{Sohn2019}%
  \BibitemOpen
  \bibfield  {author} {\bibinfo {author} {\bibfnamefont {H.~R.~O.}\
  \bibnamefont {Sohn}}, \bibinfo {author} {\bibfnamefont {S.~M.}\ \bibnamefont
  {Vlasov}}, \bibinfo {author} {\bibfnamefont {V.~M.}\ \bibnamefont {Uzdin}},
  \bibinfo {author} {\bibfnamefont {A.~O.}\ \bibnamefont {Leonov}},\ and\
  \bibinfo {author} {\bibfnamefont {I.~I.}\ \bibnamefont {Smalyukh}},\ }\href
  {https://doi.org/10.1103/PhysRevB.100.104401} {\bibfield  {journal} {\bibinfo
   {journal} {Phys. Rev. B}\ }\textbf {\bibinfo {volume} {100}},\ \bibinfo
  {pages} {104401} (\bibinfo {year} {2019})}\BibitemShut {NoStop}%
\bibitem [{\citenamefont {Leonov}\ \emph {et~al.}(2021)\citenamefont {Leonov},
  \citenamefont {Pappas},\ and\ \citenamefont {Smalyukh}}]{Leonov21}%
  \BibitemOpen
  \bibfield  {author} {\bibinfo {author} {\bibfnamefont {A.~O.}\ \bibnamefont
  {Leonov}}, \bibinfo {author} {\bibfnamefont {C.}~\bibnamefont {Pappas}},\
  and\ \bibinfo {author} {\bibfnamefont {I.~I.}\ \bibnamefont {Smalyukh}},\
  }\href@noop {} {\  (\bibinfo {year} {2021})},\ \Eprint
  {https://arxiv.org/abs/arXiv:2103.12950} {arXiv:2103.12950} \BibitemShut
  {NoStop}%
\bibitem [{\citenamefont {M\"uller}\ \emph {et~al.}(2020)\citenamefont
  {M\"uller}, \citenamefont {Rybakov}, \citenamefont {J\'onsson}, \citenamefont
  {Bl\"ugel},\ and\ \citenamefont {Kiselev}}]{Mueller2020}%
  \BibitemOpen
  \bibfield  {author} {\bibinfo {author} {\bibfnamefont {G.~P.}\ \bibnamefont
  {M\"uller}}, \bibinfo {author} {\bibfnamefont {F.~N.}\ \bibnamefont
  {Rybakov}}, \bibinfo {author} {\bibfnamefont {H.}~\bibnamefont {J\'onsson}},
  \bibinfo {author} {\bibfnamefont {S.}~\bibnamefont {Bl\"ugel}},\ and\
  \bibinfo {author} {\bibfnamefont {N.~S.}\ \bibnamefont {Kiselev}},\ }\href
  {https://doi.org/10.1103/PhysRevB.101.184405} {\bibfield  {journal} {\bibinfo
   {journal} {Phys. Rev. B}\ }\textbf {\bibinfo {volume} {101}},\ \bibinfo
  {pages} {184405} (\bibinfo {year} {2020})}\BibitemShut {NoStop}%
\bibitem [{\citenamefont {Voinescu}\ \emph {et~al.}(2020)\citenamefont
  {Voinescu}, \citenamefont {Tai},\ and\ \citenamefont
  {Smalyukh}}]{Voinescu2020}%
  \BibitemOpen
  \bibfield  {author} {\bibinfo {author} {\bibfnamefont {R.}~\bibnamefont
  {Voinescu}}, \bibinfo {author} {\bibfnamefont {J.-S.~B.}\ \bibnamefont
  {Tai}},\ and\ \bibinfo {author} {\bibfnamefont {I.~I.}\ \bibnamefont
  {Smalyukh}},\ }\href {https://doi.org/10.1103/PhysRevLett.125.057201}
  {\bibfield  {journal} {\bibinfo  {journal} {Phys. Rev. Lett.}\ }\textbf
  {\bibinfo {volume} {125}},\ \bibinfo {pages} {057201} (\bibinfo {year}
  {2020})}\BibitemShut {NoStop}%
\bibitem [{Note1()}]{Note1}%
  \BibitemOpen
  \bibinfo {note} {The displacement field $\protect \bm {u}$ describes small
  deviations of the fronts of constant phases $\phi (\protect \bm {r})=\protect
  \text {const}$ of the helimagnetic order where $\phi (\protect \bm {r})=2\pi
  (\protect \bm {r}+\protect \bm {u}) \protect \hat {\protect \bm {z}}/\lambda
  _h$.}\BibitemShut {Stop}%
\bibitem [{SI()}]{SI}%
  \BibitemOpen
  \href@noop {} {\bibinfo {title} {Supplementary information}}\BibitemShut
  {NoStop}%
\bibitem [{\citenamefont {Santangelo}(2006)}]{Santangelo06}%
  \BibitemOpen
  \bibfield  {author} {\bibinfo {author} {\bibfnamefont {C.~D.}\ \bibnamefont
  {Santangelo}},\ }\href {https://doi.org/10.1080/14645180601168117} {\bibfield
   {journal} {\bibinfo  {journal} {Liquid Crystals Today}\ }\textbf {\bibinfo
  {volume} {15}},\ \bibinfo {pages} {11} (\bibinfo {year} {2006})}\BibitemShut
  {NoStop}%
\bibitem [{\citenamefont {Giamarchi}\ \emph {et~al.}(2008)\citenamefont
  {Giamarchi}, \citenamefont {R\"uegg},\ and\ \citenamefont
  {Tchernyshyov}}]{Giamarchi2008}%
  \BibitemOpen
  \bibfield  {author} {\bibinfo {author} {\bibfnamefont {T.}~\bibnamefont
  {Giamarchi}}, \bibinfo {author} {\bibfnamefont {C.}~\bibnamefont {R\"uegg}},\
  and\ \bibinfo {author} {\bibfnamefont {O.}~\bibnamefont {Tchernyshyov}},\
  }\href {https://doi.org/10.1038/nphys893} {\bibfield  {journal} {\bibinfo
  {journal} {Nature Physics}\ }\textbf {\bibinfo {volume} {4}},\ \bibinfo
  {pages} {198} (\bibinfo {year} {2008})}\BibitemShut {NoStop}%
\bibitem [{\citenamefont {Ivanov}\ and\ \citenamefont
  {Sheka}(1995)}]{Ivanov95b}%
  \BibitemOpen
  \bibfield  {author} {\bibinfo {author} {\bibfnamefont {B.~A.}\ \bibnamefont
  {Ivanov}}\ and\ \bibinfo {author} {\bibfnamefont {D.~D.}\ \bibnamefont
  {Sheka}},\ }\href {http://link.aip.org/link/?LTP/21/881/1} {\bibfield
  {journal} {\bibinfo  {journal} {Low Temperature Physics}\ }\textbf {\bibinfo
  {volume} {21}},\ \bibinfo {pages} {881} (\bibinfo {year} {1995})}\BibitemShut
  {NoStop}%
\bibitem [{Note2()}]{Note2}%
  \BibitemOpen
  \bibinfo {note} {The topological charge $c$ of a Bloch point is defined by $c
  = \DOTSI \intop \ilimits@ _V \protect \mathrm {d}\protect \bm {r} \protect
  \bm {\nabla } \cdot \protect \bm {\Omega }$ where $V$ is a small volume
  around the Bloch point and $\Omega _i = \varepsilon _{ijk} \protect \frac
  {1}{8\pi } \protect \bm {n} (\partial _j \protect \bm {n} \times \partial _k
  \protect \bm {n})$.}\BibitemShut {Stop}%
\bibitem [{\citenamefont {Vansteenkiste}\ \emph {et~al.}(2014)\citenamefont
  {Vansteenkiste}, \citenamefont {Leliaert}, \citenamefont {Dvornik},
  \citenamefont {Helsen}, \citenamefont {Garcia-Sanchez},\ and\ \citenamefont
  {Van~Waeyenberge}}]{Vansteenkiste14}%
  \BibitemOpen
  \bibfield  {author} {\bibinfo {author} {\bibfnamefont {A.}~\bibnamefont
  {Vansteenkiste}}, \bibinfo {author} {\bibfnamefont {J.}~\bibnamefont
  {Leliaert}}, \bibinfo {author} {\bibfnamefont {M.}~\bibnamefont {Dvornik}},
  \bibinfo {author} {\bibfnamefont {M.}~\bibnamefont {Helsen}}, \bibinfo
  {author} {\bibfnamefont {F.}~\bibnamefont {Garcia-Sanchez}},\ and\ \bibinfo
  {author} {\bibfnamefont {B.}~\bibnamefont {Van~Waeyenberge}},\ }\href
  {https://doi.org/10.1063/1.4899186} {\bibfield  {journal} {\bibinfo
  {journal} {AIP Advances}\ }\textbf {\bibinfo {volume} {4}},\ \bibinfo {pages}
  {107133} (\bibinfo {year} {2014})}\BibitemShut {NoStop}%
\bibitem [{OOM()}]{OOMMFa}%
  \BibitemOpen
  \href {http://math.nist.gov/oommf/} {\bibinfo {title} {The {O}bject
  {O}riented {M}icro{M}agnetic {F}ramework}},\ \bibinfo {note} {developed by M.
  J. Donahue and D. Porter mainly, from NIST. We used the 3D version of the
  1.2b3 release}\BibitemShut {NoStop}%
\bibitem [{\citenamefont {Cort{\'e}s-Ortu{\~n}o}\ \emph
  {et~al.}(2018)\citenamefont {Cort{\'e}s-Ortu{\~n}o}, \citenamefont {Beg},
  \citenamefont {Nehruji}, \citenamefont {Pepper},\ and\ \citenamefont
  {Fangohr}}]{Cortes-Ortuno18}%
  \BibitemOpen
  \bibfield  {author} {\bibinfo {author} {\bibfnamefont {D.}~\bibnamefont
  {Cort{\'e}s-Ortu{\~n}o}}, \bibinfo {author} {\bibfnamefont {M.}~\bibnamefont
  {Beg}}, \bibinfo {author} {\bibfnamefont {V.}~\bibnamefont {Nehruji}},
  \bibinfo {author} {\bibfnamefont {R.~A.}\ \bibnamefont {Pepper}},\ and\
  \bibinfo {author} {\bibfnamefont {H.}~\bibnamefont {Fangohr}},\ }\href
  {https://doi.org/10.5281/zenodo.1196820} {\bibinfo {title} {{OOMMF}
  extension: {D}zyaloshinskii-{M}oriya interaction ({DMI}) for crystallographic
  classes {T} and {O}}} (\bibinfo {year} {2018})\BibitemShut {NoStop}%
\bibitem [{\citenamefont {Karhu}\ \emph {et~al.}(2012)\citenamefont {Karhu},
  \citenamefont {R{\"{o}}{\ss}ler}, \citenamefont {Bogdanov}, \citenamefont
  {Kahwaji}, \citenamefont {Kirby}, \citenamefont {Fritzsche}, \citenamefont
  {Robertson}, \citenamefont {Majkrzak},\ and\ \citenamefont
  {Monchesky}}]{Karhu12}%
  \BibitemOpen
  \bibfield  {author} {\bibinfo {author} {\bibfnamefont {E.~A.}\ \bibnamefont
  {Karhu}}, \bibinfo {author} {\bibfnamefont {U.~K.}\ \bibnamefont
  {R{\"{o}}{\ss}ler}}, \bibinfo {author} {\bibfnamefont {A.~N.}\ \bibnamefont
  {Bogdanov}}, \bibinfo {author} {\bibfnamefont {S.}~\bibnamefont {Kahwaji}},
  \bibinfo {author} {\bibfnamefont {B.~J.}\ \bibnamefont {Kirby}}, \bibinfo
  {author} {\bibfnamefont {H.}~\bibnamefont {Fritzsche}}, \bibinfo {author}
  {\bibfnamefont {M.~D.}\ \bibnamefont {Robertson}}, \bibinfo {author}
  {\bibfnamefont {C.~F.}\ \bibnamefont {Majkrzak}},\ and\ \bibinfo {author}
  {\bibfnamefont {T.~L.}\ \bibnamefont {Monchesky}},\ }\href
  {https://doi.org/10.1103/physrevb.85.094429} {\bibfield  {journal} {\bibinfo
  {journal} {Physical Review B}\ }\textbf {\bibinfo {volume} {85}},\ \bibinfo
  {pages} {094429} (\bibinfo {year} {2012})}\BibitemShut {NoStop}%
\bibitem [{\citenamefont {Wilson}\ \emph {et~al.}(2013)\citenamefont {Wilson},
  \citenamefont {Karhu}, \citenamefont {Lake}, \citenamefont {Quigley},
  \citenamefont {Meynell}, \citenamefont {Bogdanov}, \citenamefont {Fritzsche},
  \citenamefont {R\"o\ss{}ler},\ and\ \citenamefont {Monchesky}}]{Wilson2013}%
  \BibitemOpen
  \bibfield  {author} {\bibinfo {author} {\bibfnamefont {M.~N.}\ \bibnamefont
  {Wilson}}, \bibinfo {author} {\bibfnamefont {E.~A.}\ \bibnamefont {Karhu}},
  \bibinfo {author} {\bibfnamefont {D.~P.}\ \bibnamefont {Lake}}, \bibinfo
  {author} {\bibfnamefont {A.~S.}\ \bibnamefont {Quigley}}, \bibinfo {author}
  {\bibfnamefont {S.}~\bibnamefont {Meynell}}, \bibinfo {author} {\bibfnamefont
  {A.~N.}\ \bibnamefont {Bogdanov}}, \bibinfo {author} {\bibfnamefont
  {H.}~\bibnamefont {Fritzsche}}, \bibinfo {author} {\bibfnamefont {U.~K.}\
  \bibnamefont {R\"o\ss{}ler}},\ and\ \bibinfo {author} {\bibfnamefont {T.~L.}\
  \bibnamefont {Monchesky}},\ }\href
  {https://doi.org/10.1103/PhysRevB.88.214420} {\bibfield  {journal} {\bibinfo
  {journal} {Phys. Rev. B}\ }\textbf {\bibinfo {volume} {88}},\ \bibinfo
  {pages} {214420} (\bibinfo {year} {2013})}\BibitemShut {NoStop}%
\bibitem [{\citenamefont {Rohart}\ and\ \citenamefont
  {Thiaville}(2013)}]{Rohart2013}%
  \BibitemOpen
  \bibfield  {author} {\bibinfo {author} {\bibfnamefont {S.}~\bibnamefont
  {Rohart}}\ and\ \bibinfo {author} {\bibfnamefont {A.}~\bibnamefont
  {Thiaville}},\ }\href {https://doi.org/10.1103/PhysRevB.88.184422} {\bibfield
   {journal} {\bibinfo  {journal} {Phys. Rev. B}\ }\textbf {\bibinfo {volume}
  {88}},\ \bibinfo {pages} {184422} (\bibinfo {year} {2013})}\BibitemShut
  {NoStop}%
\bibitem [{\citenamefont {Dzyaloshinskii}\ and\ \citenamefont
  {Ivanov}(1979)}]{Dzyaloshinskii79}%
  \BibitemOpen
  \bibfield  {author} {\bibinfo {author} {\bibfnamefont {I.~E.}\ \bibnamefont
  {Dzyaloshinskii}}\ and\ \bibinfo {author} {\bibfnamefont {B.~A.}\
  \bibnamefont {Ivanov}},\ }\href@noop {} {\bibfield  {journal} {\bibinfo
  {journal} {JETP Letters}\ }\textbf {\bibinfo {volume} {29}},\ \bibinfo
  {pages} {592} (\bibinfo {year} {1979})}\BibitemShut {NoStop}%
\bibitem [{\citenamefont {Kosevich}\ \emph {et~al.}(1990)\citenamefont
  {Kosevich}, \citenamefont {Ivanov},\ and\ \citenamefont
  {Kovalev}}]{Kosevich90}%
  \BibitemOpen
  \bibfield  {author} {\bibinfo {author} {\bibfnamefont {A.~M.}\ \bibnamefont
  {Kosevich}}, \bibinfo {author} {\bibfnamefont {B.~A.}\ \bibnamefont
  {Ivanov}},\ and\ \bibinfo {author} {\bibfnamefont {A.~S.}\ \bibnamefont
  {Kovalev}},\ }\href {https://doi.org/10.1016/0370-1573(90)90130-T} {\bibfield
   {journal} {\bibinfo  {journal} {Physics Reports}\ }\textbf {\bibinfo
  {volume} {194}},\ \bibinfo {pages} {117} (\bibinfo {year}
  {1990})}\BibitemShut {NoStop}%
\bibitem [{\citenamefont {Ackerman}\ and\ \citenamefont
  {Smalyukh}(2017)}]{Ackerman17}%
  \BibitemOpen
  \bibfield  {author} {\bibinfo {author} {\bibfnamefont {P.~J.}\ \bibnamefont
  {Ackerman}}\ and\ \bibinfo {author} {\bibfnamefont {I.~I.}\ \bibnamefont
  {Smalyukh}},\ }\href {https://doi.org/10.1103/physrevx.7.011006} {\bibfield
  {journal} {\bibinfo  {journal} {Physical Review X}\ }\textbf {\bibinfo
  {volume} {7}},\ \bibinfo {pages} {011006} (\bibinfo {year}
  {2017})}\BibitemShut {NoStop}%
\bibitem [{\citenamefont {Tai}\ and\ \citenamefont {Smalyukh}(2018)}]{Tai18}%
  \BibitemOpen
  \bibfield  {author} {\bibinfo {author} {\bibfnamefont {J.-S.~B.}\
  \bibnamefont {Tai}}\ and\ \bibinfo {author} {\bibfnamefont {I.~I.}\
  \bibnamefont {Smalyukh}},\ }\href
  {https://doi.org/10.1103/physrevlett.121.187201} {\bibfield  {journal}
  {\bibinfo  {journal} {Physical Review Letters}\ }\textbf {\bibinfo {volume}
  {121}},\ \bibinfo {pages} {187201} (\bibinfo {year} {2018})}\BibitemShut
  {NoStop}%
\bibitem [{\citenamefont {Sutcliffe}(2018)}]{Sutcliffe18}%
  \BibitemOpen
  \bibfield  {author} {\bibinfo {author} {\bibfnamefont {P.}~\bibnamefont
  {Sutcliffe}},\ }\href {https://doi.org/10.1088/1751-8121/aad521} {\bibfield
  {journal} {\bibinfo  {journal} {Journal of Physics A: Mathematical and
  Theoretical}\ }\textbf {\bibinfo {volume} {51}},\ \bibinfo {pages} {375401}
  (\bibinfo {year} {2018})}\BibitemShut {NoStop}%
\bibitem [{\citenamefont {Hopf}(1931)}]{Hopf31}%
  \BibitemOpen
  \bibfield  {author} {\bibinfo {author} {\bibfnamefont {H.}~\bibnamefont
  {Hopf}},\ }\href {https://doi.org/10.1007/bf01457962} {\bibfield  {journal}
  {\bibinfo  {journal} {Mathematische Annalen}\ }\textbf {\bibinfo {volume}
  {104}},\ \bibinfo {pages} {637} (\bibinfo {year} {1931})}\BibitemShut
  {NoStop}%
\bibitem [{\citenamefont {Kravchuk}\ and\ \citenamefont
  {Sheka}(2007)}]{Kravchuk07a}%
  \BibitemOpen
  \bibfield  {author} {\bibinfo {author} {\bibfnamefont {V.}~\bibnamefont
  {Kravchuk}}\ and\ \bibinfo {author} {\bibfnamefont {D.}~\bibnamefont
  {Sheka}},\ }\href {http://dx.doi.org/10.1134/S1063783407100186} {\bibfield
  {journal} {\bibinfo  {journal} {Physics of the Solid State}\ }\textbf
  {\bibinfo {volume} {49}},\ \bibinfo {pages} {1923} (\bibinfo {year}
  {2007})}\BibitemShut {NoStop}%
\bibitem [{\citenamefont {Malozemoff}\ and\ \citenamefont
  {Slonzewski}(1979)}]{Malozemoff79}%
  \BibitemOpen
  \bibfield  {author} {\bibinfo {author} {\bibfnamefont {A.~P.}\ \bibnamefont
  {Malozemoff}}\ and\ \bibinfo {author} {\bibfnamefont {J.~C.}\ \bibnamefont
  {Slonzewski}},\ }\href@noop {} {\emph {\bibinfo {title} {Magnetic domain
  walls in bubble materials}}}\ (\bibinfo  {publisher} {Academic Press},\
  \bibinfo {address} {New York},\ \bibinfo {year} {1979})\BibitemShut {NoStop}%
\bibitem [{\citenamefont {D\"{o}ring}(1968)}]{Doering68}%
  \BibitemOpen
  \bibfield  {author} {\bibinfo {author} {\bibfnamefont {W.}~\bibnamefont
  {D\"{o}ring}},\ }\href {https://doi.org/10.1063/1.1656144} {\bibfield
  {journal} {\bibinfo  {journal} {Journal of Applied Physics}\ }\textbf
  {\bibinfo {volume} {39}},\ \bibinfo {pages} {1006} (\bibinfo {year}
  {1968})}\BibitemShut {NoStop}%
\bibitem [{\citenamefont {Pylypovskyi}\ \emph {et~al.}(2012)\citenamefont
  {Pylypovskyi}, \citenamefont {Sheka},\ and\ \citenamefont
  {Gaididei}}]{Pylypovskyi12}%
  \BibitemOpen
  \bibfield  {author} {\bibinfo {author} {\bibfnamefont {O.~V.}\ \bibnamefont
  {Pylypovskyi}}, \bibinfo {author} {\bibfnamefont {D.~D.}\ \bibnamefont
  {Sheka}},\ and\ \bibinfo {author} {\bibfnamefont {Y.}~\bibnamefont
  {Gaididei}},\ }\href {https://doi.org/10.1103/PhysRevB.85.224401} {\bibfield
  {journal} {\bibinfo  {journal} {Physical Review B}\ }\textbf {\bibinfo
  {volume} {85}},\ \bibinfo {pages} {224401} (\bibinfo {year}
  {2012})}\BibitemShut {NoStop}%
\bibitem [{\citenamefont {Charilaou}(2020)}]{Charilaou20}%
  \BibitemOpen
  \bibfield  {author} {\bibinfo {author} {\bibfnamefont {M.}~\bibnamefont
  {Charilaou}},\ }\href {https://doi.org/10.1103/physrevb.102.014430}
  {\bibfield  {journal} {\bibinfo  {journal} {Physical Review B}\ }\textbf
  {\bibinfo {volume} {102}},\ \bibinfo {pages} {014430} (\bibinfo {year}
  {2020})}\BibitemShut {NoStop}%
\bibitem [{\citenamefont {Li}\ \emph {et~al.}(2020)\citenamefont {Li},
  \citenamefont {Pierobon}, \citenamefont {Charilaou}, \citenamefont {Braun},
  \citenamefont {Walet}, \citenamefont {Löffler}, \citenamefont {Miles},\ and\
  \citenamefont {Moutafis}}]{Li20b}%
  \BibitemOpen
  \bibfield  {author} {\bibinfo {author} {\bibfnamefont {Y.}~\bibnamefont
  {Li}}, \bibinfo {author} {\bibfnamefont {L.}~\bibnamefont {Pierobon}},
  \bibinfo {author} {\bibfnamefont {M.}~\bibnamefont {Charilaou}}, \bibinfo
  {author} {\bibfnamefont {H.-B.}\ \bibnamefont {Braun}}, \bibinfo {author}
  {\bibfnamefont {N.~R.}\ \bibnamefont {Walet}}, \bibinfo {author}
  {\bibfnamefont {J.~F.}\ \bibnamefont {Löffler}}, \bibinfo {author}
  {\bibfnamefont {J.~J.}\ \bibnamefont {Miles}},\ and\ \bibinfo {author}
  {\bibfnamefont {C.}~\bibnamefont {Moutafis}},\ }\href
  {https://doi.org/10.1103/physrevresearch.2.033006} {\bibfield  {journal}
  {\bibinfo  {journal} {Physical Review Research}\ }\textbf {\bibinfo {volume}
  {2}},\ \bibinfo {pages} {033006} (\bibinfo {year} {2020})}\BibitemShut
  {NoStop}%
\end{thebibliography}%


%

\clearpage
\break
\onecolumngrid






\begin{center}
	\textbf{\large SUPPLEMENTARY MATERIALS}
\end{center}
	\beginsupplement
	
		In section I we provide details of the micromagnetic simulations. In section II we present the derivation of the universal far field configuration of screw dislocations in chiral magnets and compare with micromagnetic simulations. In addition, we discuss the influence of rotational symmetry breaking on the screw dislocation energy at zero field. 
		In section III the micromagnetic core structures of screw dislocations is further analyzed. In particular, we demonstrate the use of linking numbers to distinguish between the configurations sd$_1^{\text{Sk}\pm}$ and sd$_1^\pm$, and we further compare the skyrmion tube Sk$^+$ with the screw dislocation sd$_1^{\text{Sk}+}$. Finally, we discuss the energy and positions of Bloch points contained in the screw dislocation sd$^{\rm Bp}_1$.
	%

	
	\section{Micromagnetic simulations of screw dislocations in chiral magnets}\label{app:Vd12}
	
	For the micromagnetic simulations we used both MuMax \cite{Vansteenkiste14} and OOMMF \cite{OOMMFa,Cortes-Ortuno18} with a discrete cubic lattice with an elementary unit cell with size $\Delta x=\Delta y=\Delta z=0.5$ nm. We simulated a cylindrical sample with periodic boundary conditions along the cylinder axis, i.e., $z$-axis that coincides with the magnetic field direction. On the edge of the cylinder we applied free boundary conditions. For concreteness, we used parameters appropriate fro MnSi, i.e., $A=0.4$~pJ/m, $\mu_0M_s=0.203$~T, $D=0.28$~mJ/m$^2$ \cite{Karhu12}, implying 
	$\lambda_h\approx18$ nm and $\mu_0H_{c2}=0.6$ T. The  size of the simulated volume was $L_z=m\lambda_h$ along the $z$-axis with an integer $m$ between $1<m\le4$. The radius of the cylinder was varied between 5 and 11 times the helix wavelength $\lambda_h$.
	
	\section{Far field of screw dislocations in chiral magnets}\label{app:screw-asympt}
	
	\subsection{Asymptotic expansion of the Euler-Lagrange equations}
	
	The Euler-Lagrange equations for the energy density functional of Eq. (2) in the main text can be written in the form 
	\begin{subequations}\label{eq:angular-stat}
		\begin{align}
			\label{eq:theta}	
			-\vec{\nabla}^2\theta+\sin\theta\cos\theta\left[(\vec{\nabla}\phi)^2-2\partial_\zeta\phi\right]+h\sin\theta
			-2\sin^2\theta(\vec{\varepsilon}\cdot\vec{\nabla}\phi)&=0,
			\\\label{eq:phi}
			-\vec{\nabla}\cdot\left(\sin^2\theta\vec{\nabla}\phi\right)+2\sin^2\theta\left(\vec{\nabla}\theta\cdot\vec{\varepsilon}\right)+2\sin\theta\cos\theta\partial_{\zeta}\theta&=0.
		\end{align}	
	\end{subequations}
	Here, we used the angular parameterization of the unit vector 
	$\vec{n}=\sin\theta\vec{\varepsilon}+\cos\theta\hat{\vec{z}}$ with $\vec{\varepsilon}=\hat{\vec{x}}\cos\phi+\hat{\vec{y}}\sin\phi$, and the derivatives are taken with respect to the dimensionless coordinate $2\pi\vec{r}/\lambda_h$ with $\zeta=2\pi z/\lambda_h$ being the dimensionless coordinate in direction of the field $h = H/H_{c2}$. Here, we denote the dimensionless cylindrical coordinates by $\{\rho,\chi,\zeta\}$. 
	
	Let us first consider the case $h=0$ where the ground state is given by the helix with $\theta=\pi/2$, $\phi=\zeta$. Substituting into \eqref{eq:angular-stat} a vortex-like solution $\theta =\pi/2$, $\phi =\nu\chi+\zeta$, one can see that Eq.~\eqref{eq:phi} is fulfilled but Eq.~\eqref{eq:theta} has a non-vanishing left-hand-side (LHS) of order $\mathcal{O}(\rho^{-1})$. 
	In order to compensate this contribution, we look instead for solutions of the form $\theta=\pi/2+\vartheta_1(\chi,\zeta)/\rho$ and $\phi=\nu\chi+\zeta$. In the limit of large $\rho$  Eq.~\eqref{eq:theta} reduces in this case to 
	\begin{equation}\label{eq:angular-stat1}
	\frac{\partial_{\zeta\zeta}^2\vartheta_1-\vartheta_1+2\nu\sin[(\nu-1)\chi+\zeta]}{\rho}+\mathcal{O}\left(\frac{1}{\rho^3}\right)=0,
	\end{equation}
	and the LHS of Eq.~\eqref{eq:phi} is only of order $\mathcal{O}\left(\rho^{-2}\right)$. At this order, the equations are fulfilled by choosing $\vartheta_1=\nu\sin[(\nu-1)\chi+\zeta]$ so that the numerator in \eqref{eq:angular-stat1} vanishes. The accuracy can be iteratively improved in a next step by looking for a solution with the form $\theta=\pi/2+\vartheta_1/\rho$ and $\phi=\nu\chi+\zeta+\varphi_2(\chi,\zeta)/\rho^2$, where $\vartheta_1$ was found in the previous step. The 
	LHS of Eq.~\eqref{eq:theta} is still on the order $\mathcal{O}\left(\rho^{-3}\right)$ and  
	Eq.~\eqref{eq:phi} obtains the form
	\begin{equation}\label{eq:angular-stat2}	
	\frac{\partial_{\zeta\zeta}^2\varphi_2+\nu(2-\nu)\sin[2(\nu-1)\chi+2\zeta]}{\rho^2}+\mathcal{O}\left(\frac{1}{\rho^4}\right)=0.
	\end{equation}
	The function $\varphi_2$ is determined by demanding that this equation vanishes up to order $\mathcal{O}\left(\frac{1}{\rho^4}\right)$ yielding $\varphi_2=\frac14\nu(2-\nu)\sin[2(\nu-1)\chi+2\zeta]$.

	Repeating this procedure we obtain the asymptotic solution for the screw dislocation in the form of a series up to any required accuracy. Up to terms of order $\mathcal{O}\left(\rho^{-5}\right)$ this solution is
	\begin{subequations}\label{eq:screw}
		\begin{align}
			\label{eq:screw-theta}	&\theta=\frac{\pi}{2}+\frac{a_{11}\sin\psi}{\rho}+\frac{a_{21}\sin\psi+a_{22}\sin3\psi}{\rho^3}+\mathcal{O}\left(\frac{1}{\rho^5}\right)\\
			\label{eq:screw-phi}	&\phi=\nu\chi+\zeta+\frac{b_{11}\sin2\psi}{\rho^2}+\frac{b_{21}\sin2\psi+b_{22}\sin4\psi}{\rho^4} +\mathcal{O}\left(\frac{1}{\rho^6}\right),
		\end{align}	
	\end{subequations}
	where $\psi=(\nu-1)\chi+\zeta$ and coefficients $a_{ij}$ and $b_{ij}$ are given by
	\begin{equation}\label{eq:ab}
	\begin{split}
	&a_{11}=\nu,\quad a_{21}=\frac{3}{8}\nu^2(2-\nu),\quad a_{22}=-\nu\left(\frac15-\frac{\nu}{4}+\frac{\nu^2}{24}\right);\\
	&b_{11}=\frac{\nu(2-\nu)}{4},\quad b_{21}=\nu^2\left(\frac85-\frac{5\nu}{4}+\frac{\nu^2}{8}\right),\quad
	b_{22}=-\nu(2-\nu)\left(\frac{3}{80}-\frac{\nu}{16}+\frac{\nu^2}{32}\right).
	\end{split}
	\end{equation}
	Thus, one can conjecture that the deviation of the screw dislocation of the helimagnetic order from the vortex-like solution can be expressed in odd powers of $1/\rho$ for $\theta$ and in even powers for $\phi$.

	However, this symmetry is broken by the magnetic field when the helix is becoming conical. Applying the procedure described above for the case $h\ne0$ we obtain
	\begin{subequations}\label{eq:screw-b}
		\begin{align}
			\label{eq:theta-b}\theta=&\theta_0+\frac{2\nu\sin^2\theta_0}{1+\sin^2\theta_0}\frac{\sin\psi}{\rho}
			-\frac{\nu^2h}{\sqrt{1-h^2}}\frac{\cos^4\theta_0-2\sin^4\theta_0\sin^2\psi}{\rho^2\left(1+\sin^2\theta_0\right)^2}+\mathcal{O}\left(\frac{1}{\rho^3}\right),\\
			\phi=&\nu\chi+\zeta+\frac{\nu(2-\nu)}{2}\frac{\sin^2\theta_0}{1+\sin^2\theta_0}\frac{\sin2\psi}{\rho^2}
			-\frac{4h\nu^2}{\sqrt{1-h^2}}\frac{[1-(1+\nu)\sin^2\theta_0]\cos\psi}{\rho^3(1+\sin^2\theta_0)^2}+\mathcal{O}\left(\frac{1}{\rho^4}\right),
		\end{align}
	\end{subequations}
	where $\cos\theta_0=h$. Based on \eqref{eq:screw-b} we conclude that the asymptotic expansion \eqref{eq:screw-b} is valid for $\rho\gg\max\{1,|\nu|h/\sqrt{1-h^2}\}$. As the critical field is approached $h \to 1^-$, the size of the core of the screw dislocation is governed by the correlation length $\xi \sim 1/\sqrt{1-h}$ that diverges for $h \to 1^-$, see discussion in the main text.

	Generally, the asymptotic solution is obtained with the help of the Ansatz
	\begin{equation}\label{eq:as-gen}
	\theta=\theta_0+\sum\limits_{m=1}^\infty\frac{\vartheta_m(\chi,\zeta)}{\rho^m},\quad
	\phi=\nu\chi+\zeta+\sum\limits_{m=1}^\infty\frac{\varphi_m(\chi,\zeta)}{\rho^m}.
	\end{equation}
	Substituting \eqref{eq:as-gen} into \eqref{eq:angular-stat} and collecting terms at the same powers $\rho^{-m}$ one obtains a cascade of sets of equations with two equations in each set. Solving the first set ($m=1$) results in $\vartheta_1$ and $\varphi_1$. Solving the second set ($m=2$) using the solution for $\vartheta_1$ and $\varphi_1$  results in $\vartheta_2$ and $\varphi_2$. The iterations are repeated until the required accuracy is achieved. In our particular case, it turns out that the each set consists of two independent equations (this is true at least for $m\le4$). 
	
	\subsection{Comparison of the far field with micromagnetic simulations}
	
	\begin{figure}
		\includegraphics[width=0.5\columnwidth]{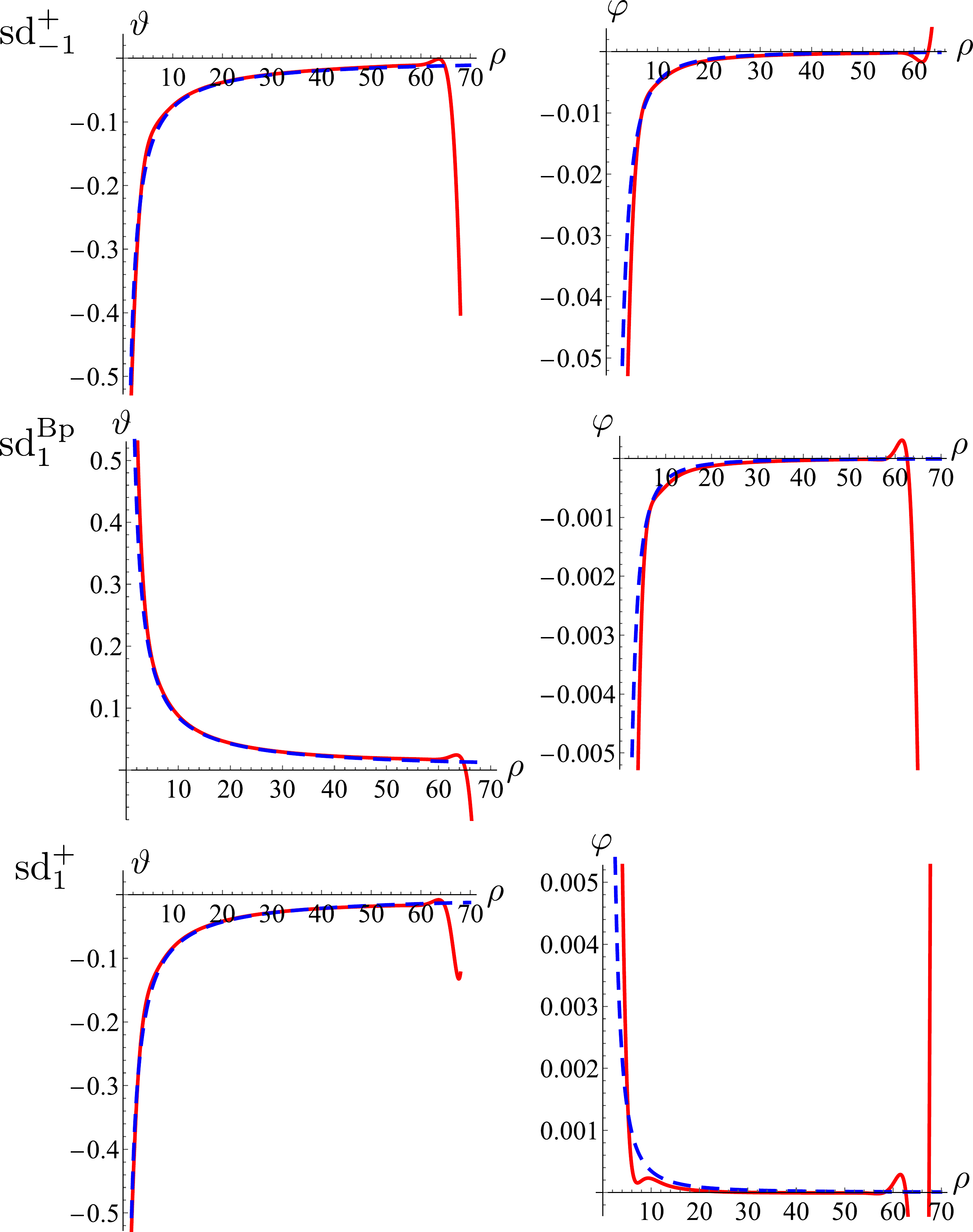}
		\caption{Far field of three distinct screw dislocations, sd$_{-1}^+$, sd$_1^{\rm Bp}$ and sd$_1^+$. The angles $\vartheta=\theta-\theta_0$ and $\varphi=\phi-\nu\chi-\zeta$ extracted from micromagnetic simulations (solid line) are compared with the asymptotic expansion of Eq.~\eqref{eq:screw-b} (dashed line). For all cases $h=0.5$, $\chi=\pi/4$, and the dimensionless coordinate $\zeta=2.7$, 1.7 and 4.1 for sd$_{-1}^+$, sd$_{1}^\text{Bp}$ and sd$_{1}^+$, respectively. The simulations were carried out for a cylinder-shaped sample with height $\lambda_h$ and radius $11.1\lambda_h$. The dimensionless radial coordinate is thus bounded between $0 < \rho < 2\pi  11.1 \approx 69.7$.
			The strong deviations close to the edge of the cylinder are due to surface twist effects. Good agreement between numerics and analytics is obtained in the intermediate range where Eq.~\eqref{eq:screw-b} is applicable, see also Fig.~\ref{fig:rhotheta}.
		}\label{fig:asympt}
	\end{figure}
	
	\begin{figure}
		\includegraphics[width=0.5\columnwidth]{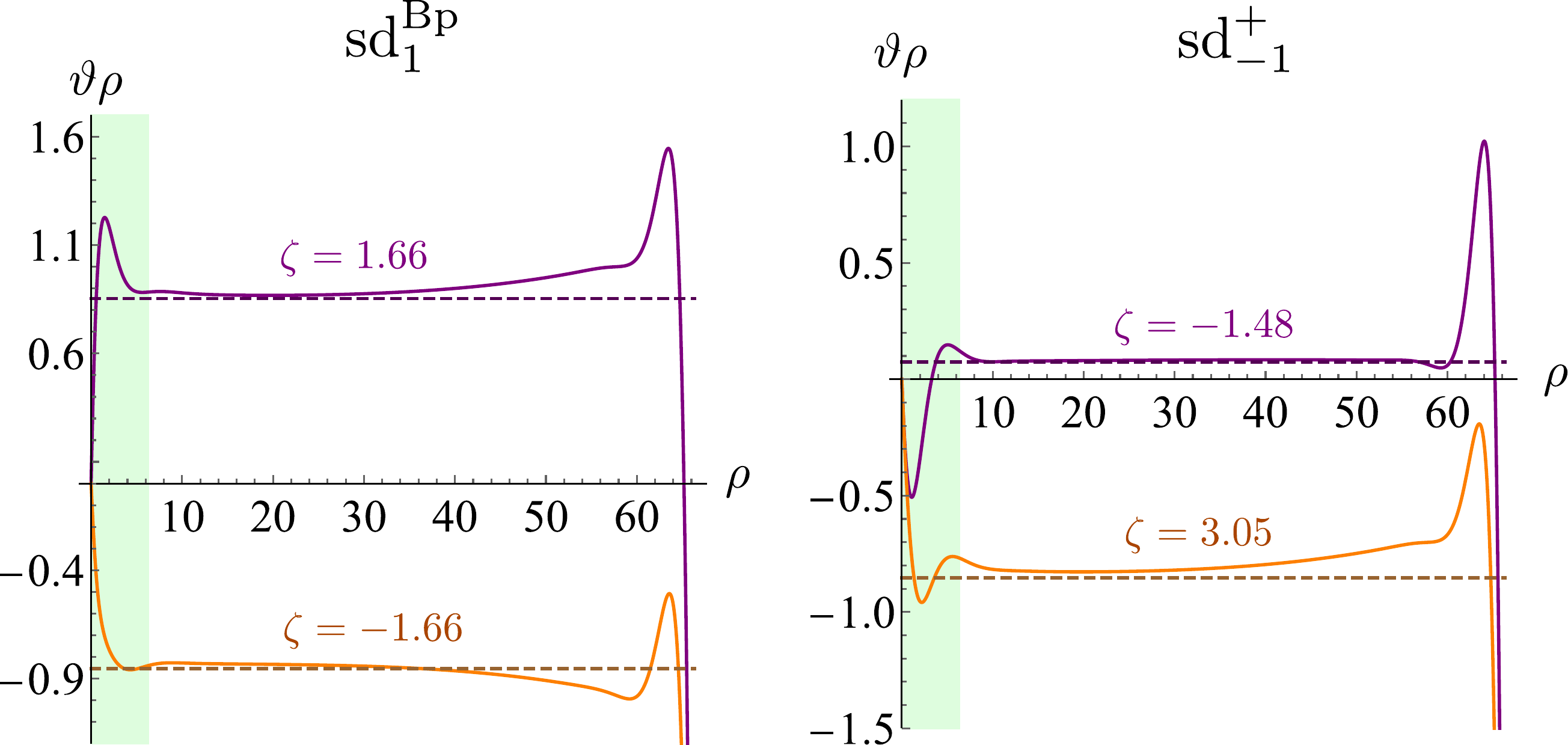}
		\caption{Further comparison of the far field between numerics and analytics for screw dislocations sd$_1^{\rm Bp}$ and sd$_{-1}^+$. The quantity $\vartheta\rho$
			obtained from the micromagnetic simulations (solid lines) is compared to its analytically expected asymptotic value $2\nu\sin^2\theta_0[1+\sin^2\theta_0]^{-1}\sin\psi$ (dashed lines) for the case $h=0.5$, $\chi=\pi/4$ for different values of $\zeta$. The green shaded area indicates the region of the dislocation core. The surface effect is quite pronounced hampering a quantitative comparison with the analytical asymptotic solution at larger distances $\rho$.  
		}
		\label{fig:rhotheta}
	\end{figure}

	The comparison of the asymptotic solutions \eqref{eq:screw-b} in the presence of a finite magnetic field with the micromagnetic simulations are presented in Fig.~\ref{fig:asympt}. Good agreement is found in the far field except close to the sample edge where the DMI-induced surface twist \cite{Wilson2013,Rohart2013} leads to significant distortions not accounted for by the analytical treatment. This is further illustrated in Fig.~\ref{fig:rhotheta} that shows $\vartheta \rho = (\theta-\theta_0) \rho$ which is expected to approach a constant value in the far field (dashed line). The approach to this value is hampered by the surface twist that appears to possess a large penetration depth in the present case; this requires further study. Fig.~\ref{fig:rhotheta} also allows to estimate the size of the screw dislocation core that is indicated by the green shaded region. For the chosen parameters in this example, the core is on the order of a single helix wavelength $\lambda_h$, i.e., $\rho \sim 2\pi$. 
	
	\subsection{Influence of uniaxial anisotropy on the screw dislocation energy}
	
	The energy per length of a screw dislocation line in general depends logarithmically on the system size, see Eq.~(6) of the main text. This logarithmic dependence disappears in zero field for the theory of Eq.~(2) of the main text where the theory possesses rotational symmetry. It was argued in the main text that a logarithmic dependence remains even at $H=0$ when magnetocrystalline anisotropies are taken into account that explicitly break the rotation symmetry. 
	
	In order to demonstrate this point, we consider in the following a modified Langrange density that contains an additional uniaxial anisotropy $K$,
	\begin{equation}
	\label{eq:H-an}
	\mathcal{E} = A (\partial_i \vec{n})^2 + D \vec{n} (\vec{\nabla} \times \vec{n}) - M_s \mu_0 H  n_z+Kn_z^2.
	\end{equation}
	Whereas an uniaxial anisotropy does not comply with the symmetry of cubic chiral magnets, it is easier to treat as compared to cubic magnetocrystalline anisotropies and it serves our purpose.
	
	One of the Euler-Lagrange equations still coincides with Eq.~\eqref{eq:phi}, whereas the other in dimensionless units is now given by 
	\begin{equation}\label{eq:angular-stat-an}
	\begin{split}
	-\vec{\nabla}^2\theta+\sin\theta\cos\theta[(\vec{\nabla}\phi)^2&-2\partial_\zeta\phi-\kappa]+h\sin\theta-2\sin^2\theta(\vec{\varepsilon}\cdot\vec{\nabla}\phi)=0,
	\end{split}	
	\end{equation}
	where $\kappa=4KA/D^2$ is the dimensionless anisotropy coefficient. The conical state solution $\phi=\zeta$ and $\theta=\theta_0$ with $\cos\theta_0=h/(1+\kappa)$ exists if $\kappa>-1$ and $|h|<1+\kappa$. For the far field of the screw dislocation we now obtain
	\begin{subequations}\label{eq:far-fld-an}
		\begin{align}
			&\theta=\theta_0+\frac{2\nu\sin^2\theta_0}{1+(1+\kappa)\sin^2\theta_0}\frac{\sin\psi}{\rho}+\mathcal{O}(\rho^{-2}),\\
			&	\phi=\nu\chi+\zeta+\frac{\nu(2-\nu)\sin^2\theta_0}{2[1+(1+\kappa)\sin^2\theta_0]}\frac{\sin2\psi}{\rho^2}+\mathcal{O}(\rho^{-3}).
		\end{align}
	\end{subequations}
	The corresponding energy per length is
	\begin{equation} \label{eq:EnergySD}
	\varepsilon_{\rm sd} = \varepsilon^{\rm core}_{\rm sd} + A (2 \pi  \nu)^2  \sin^2\theta_0\frac{1-(1-\kappa)\sin^2\theta_0}{1+(1+\kappa)\sin^2\theta_0} \log \frac{R}{\rho_{\rm core}},
	\end{equation}
	where the field dependence is incorporated into $\theta_0$. For the case $\kappa=0$ one has $\sin^2\theta_0=1-h^2$, and the result of Eq.~(6) in the main text is reproduced. For $h=0$, i.e., $\sin^2\theta_0=1$ we obtain instead
	\begin{equation} \label{eq:EnergySD-h0}
	\varepsilon_{\rm sd} = \varepsilon^{\rm core}_{\rm sd} + A (2 \pi  \nu)^2  \frac{\kappa}{2+\kappa} \log \frac{R}{\rho_{\rm core}}.
	\end{equation}
	Thus, for a non-vanishing uniaxial anisotropy $\kappa$ the screw dislocation energy is logarithmically divergent even at zero magnetic field.
	
	\section{Further discussions on the magnetic microstructure of screw dislocations}
	
	\subsection{Linking numbers for screw dislocations sd$_{1}^{\text{Sk}+}$ and sd$_1^+$}
	
	\begin{figure}[t]
		\includegraphics[width=0.5\columnwidth]{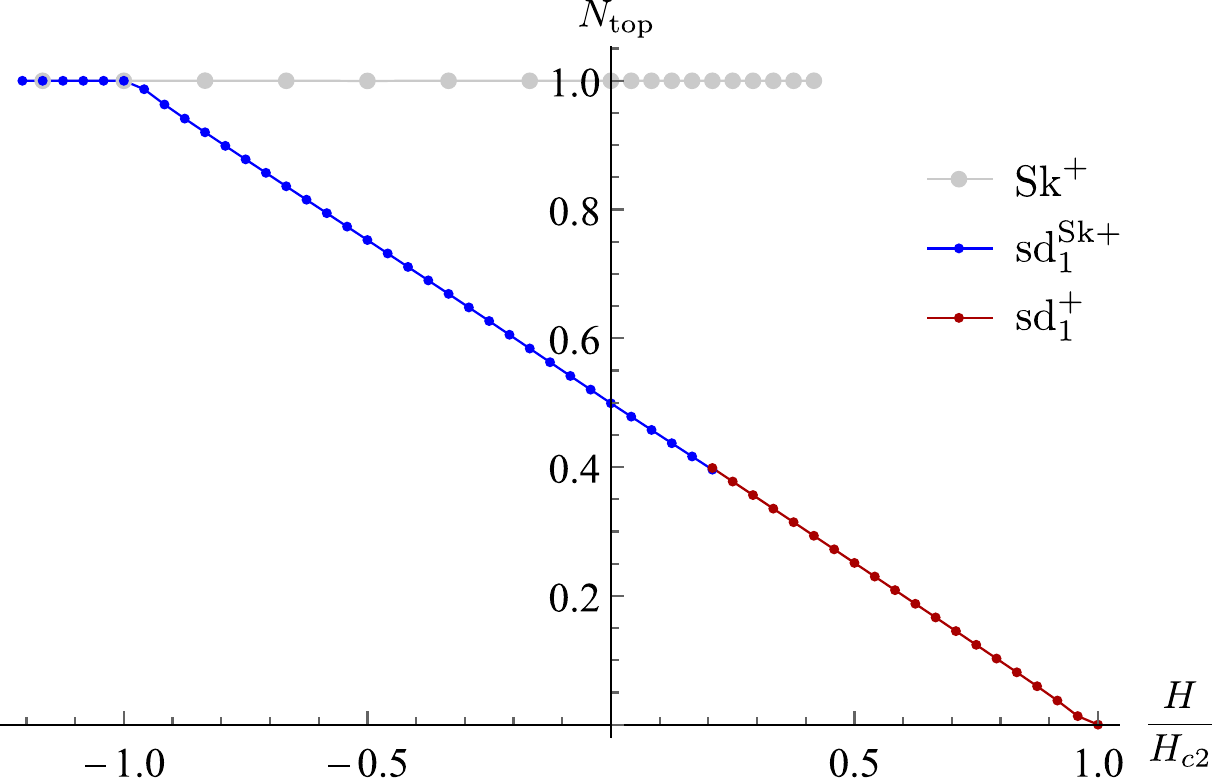}
		\caption{Field evolution of the skyrmion number  calculated for an arbitrary horizontal plane $z=\text{const}$ for screw dislocations sd$_{1}^{\text{Sk}+}$ and sd$_1^+$ as well as for the skyrmion tube Sk$^+$ as obtained from micromagnetic simulations. This explicitly confirms that the skyrmion number of both dislocations is $N_{\rm top} = \frac{1}{2}(1-H/H_{c2})$ whereas $N_{\rm top} = 1$ for Sk$^+$. Note that due to the slow algebraic decay of the dislocations' far field, an accurate determination of $N_{\text{top}}$ requires significantly larger system size within the $(x,y)$-plane as compared to the exponentially localized skyrmion tubes. Alternatively, for the particular case of $\nu=1$ one can calculate $N_{\text{top}}$ for planes $\zeta=n\pi$ (corresponding to $z=\pi\lambda_h/2$), where the leading corrections to the asymptotic solutions \eqref{eq:screw-b} vanish.
		}\label{fig:Q-vs-b}
	\end{figure}

	\begin{figure}[h!]
		\includegraphics[width=0.6\columnwidth]{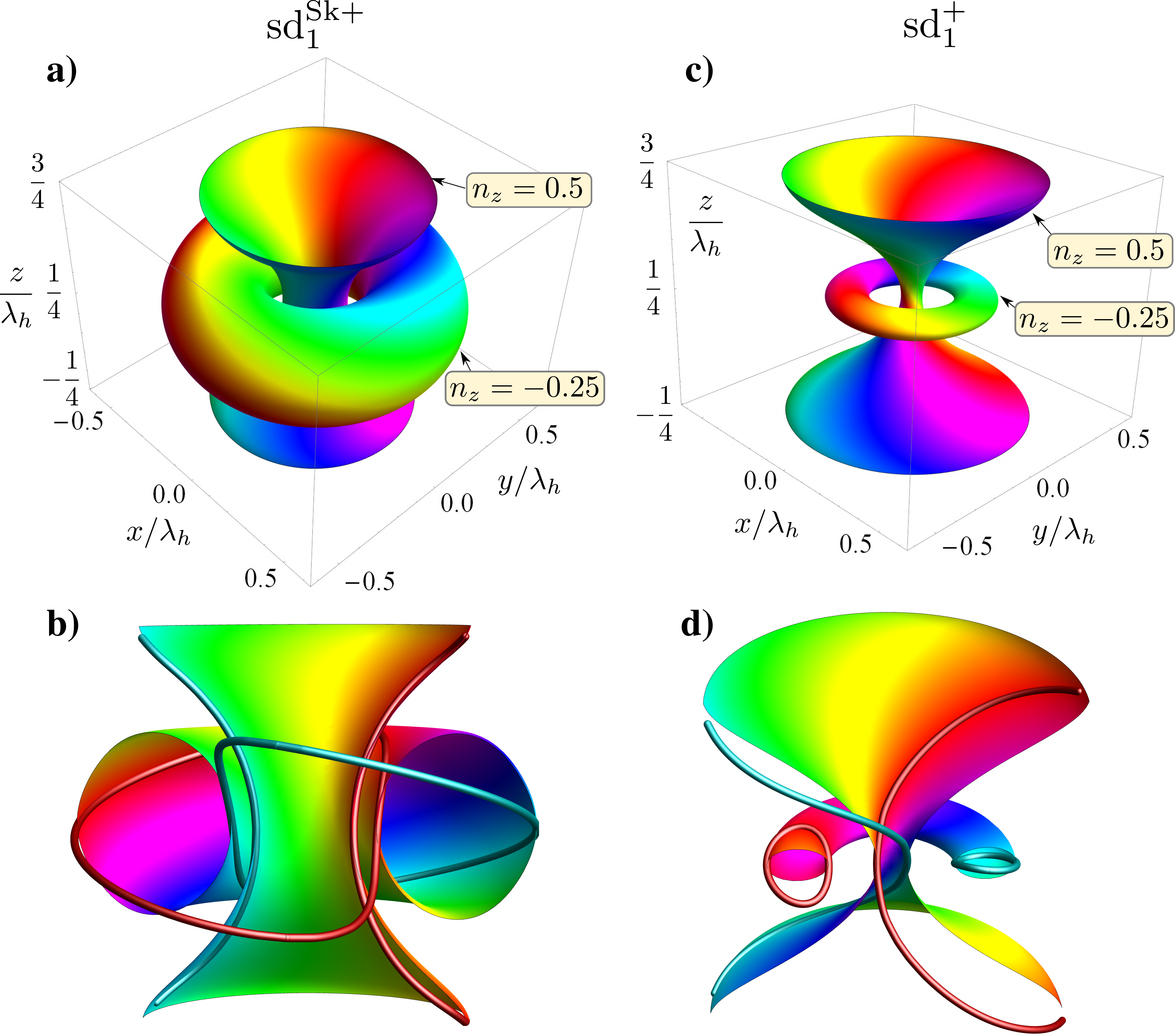}
		\caption{Isosurfaces of screw dislocations sd$_{1}^{\text{Sk}+}$ (panel (a) and (b)) and sd$_1^+$ (panel (c) and (d)) as obtained by micromagnetic simulations. The magnetic field is $H = 0.21 H_{c2}$ corresponding to points $B$ and $D$ in Fig.~3 of the main text. The coloring represents the angle $\phi=\arg(n_x+in_y)$. Panels (b) and (d) correspond to vertical cross-sections of (a) and (c), respectively, and also show isolines on the isosurfaces characterized by constants $\phi=0$ and $\phi=\pi$ corresponding to the dark-red and dark-cyan lines, respectively.
		}\label{fig:linking}
	\end{figure}
	
	Whereas the stability regime of the screw dislocations sd$_{1}^{\text{Sk}+}$ and sd$_1^+$ as a function of magnetic field is quite different, see Fig. 3 of the main text, they nevertheless share common features. They both possess the same strength $\nu= 1$, a smooth core, their core magnetization is aligned with the $z$-axis, and  the skyrmion number in each plane is for both given by $N_{\rm top} = \frac{1}{2}(1-H/H_{c2})$, see Fig.~\ref{fig:Q-vs-b} (such that sd$_{1}^{\text{Sk}+}$ smoothly connects to a skyrmion string with $N_{\rm top} = 1$ at the negative critical field $H = - H_{c2}$). Coincidentally, both become also unstable at a similar magnetic field and transform into sd$^{\rm Bp}_1$ by the nucleation of singular Bloch points, see points $B$ and $D$ in Fig.~3 of the main text. 
	
	Nevertheless, both screw dislocations sd$_{1}^{\text{Sk}+}$ and sd$_1^+$ can be distinguished topologically by  linking numbers. Both dislocations possess toroidal isosurfaces with a negative magnetization surrounding the core line, cf.~Fig.4(a) and (c) of the main text. This suggest a certain similarity with Hopfions \cite{Dzyaloshinskii79,Kosevich90,Ackerman17,Tai18,Sutcliffe18}. However in contrast to Hopfions, the screw dislocations sd$_{1}^{\text{Sk}+}$ and sd$_1^+$ are extended line defects. They are not spatially localized and, as a consequence, they can not be characterized by a Hopf invariant \cite{Hopf31}.  Note that Fig.~4 of the main text shows only a single period along the $z$-axis each containing a toroidal isosurface. These isosurfaces are again shown in Fig.~\ref{fig:linking}(a) and (c) with a coloring that is reflecting the phase $\phi=\arg(n_x+in_y)$. 
	
	The nature of the toroidal isosurfaces in both cases is quite distinct. This is best illustrated by considering the 
	plane $z = \frac{1}{4} \lambda_h$ shown in the lower panels of Fig.~4(a) and (c). When approaching the core from the far field, the magnetization first twist right-handedly for sd$_1^+$ but then reverses and twists left-handedly in order to align the magnetization with the $z$-axis at the center. In total the magnetization rotates by $\pi/2$ but due to the twist reversal a toroidal isosurface is created. In contrast, the magnetization always twists in a right-handed fashion upon approaching the core for sd$_{1}^{\text{Sk}+}$ with a total rotation angle of $3\pi/2$. The additional rotation by $\pi$ results in the toroidal isosurface for sd$_{1}^{\text{Sk}+}$. 
	
	This distinct nature of toroidal isosurfaces is reflected in the topological properties of isolines where the magnetization points in a specific direction. Examples of isolines on the toroidal isosurfaces are shown in Fig.~\ref{fig:linking}(b) and (d). Whereas pairs of isolines on the toroidal isosurfaces of sd$_{1}^{\text{Sk}+}$ are interlinked and thus characterized by a finite linking number $N_{\rm link} = 1$, the corresponding pairs for sd$_1^+$ are not linked and thus $N_{\rm link} = 0$. At the same time, pairs of isolines defined on isosurfaces close to the core are linked for sd$_1^+$ with $N_{\rm link} = 1$ and not linked for sd$_{1}^{\text{Sk}+}$ with $N_{\rm link} = 0$.
	This particular situation, that different pairs of isolines possess different linking numbers for the same screw dislocation, is here possible because these line defects are extended and not spatially localized. 
	
	\begin{figure*}[t]
		\includegraphics[width=\textwidth]{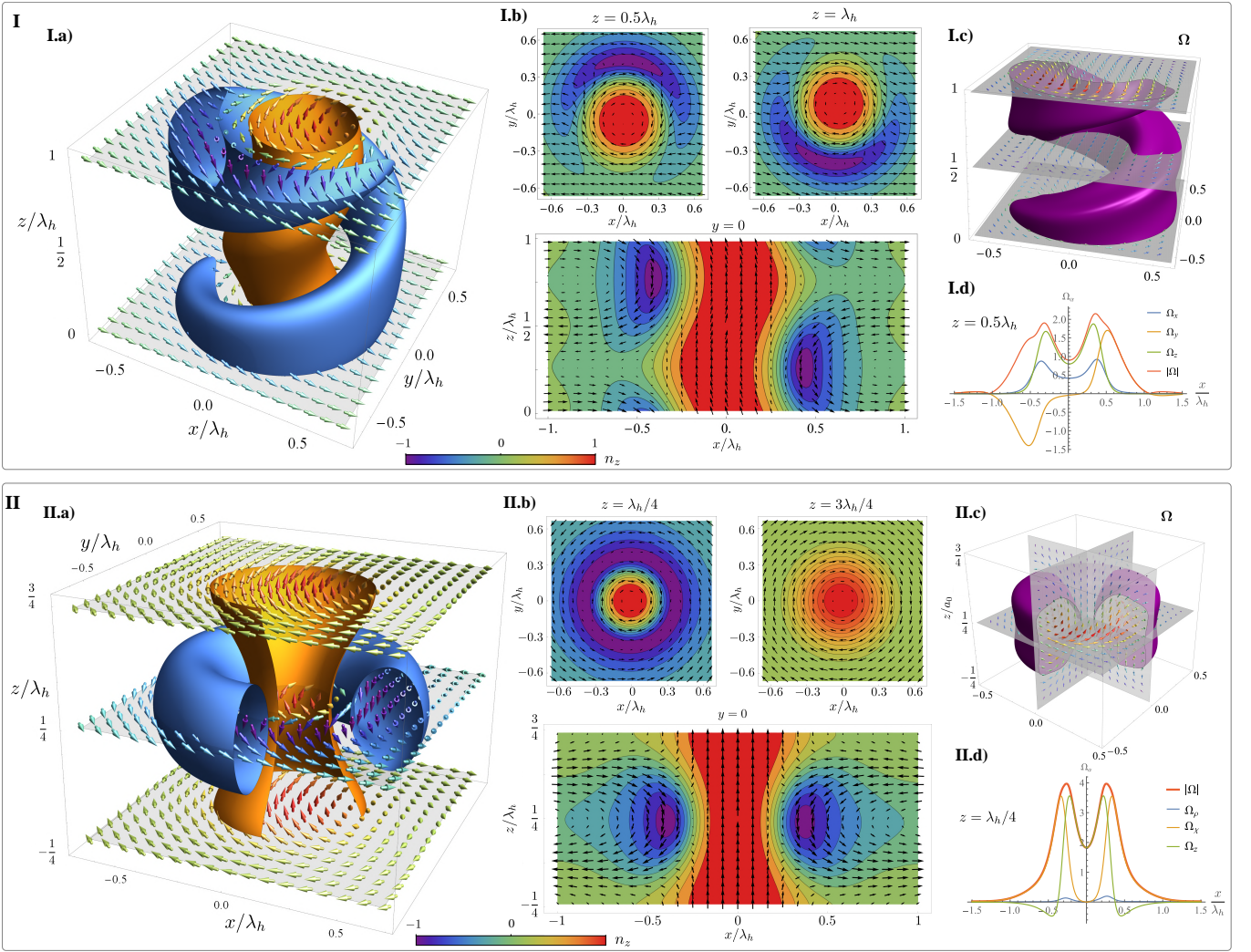}
		\caption{Comparison between the skyrmion tube Sk$^+$ in panel I and the screw dislocation sd$_1^{\text{Sk}+}$ in panel II at zero field. All data were obtained by micromagnetic simulations. Each panel (a) illustrates the micromagnetic structure within a single period $\lambda_h$ along the magnetic field axis where orange and blue isosurfaces are defined by $n_z = 1/2$ and $-1/2$, respectively. Panels (b) show cross-sections of the magnetization. Panel (c) displays the distribution of the gyrofield $\vec{\Omega}=\hat{\vec{x}}_i\frac{\epsilon_{ijk}}{8\pi}\vec{n}\cdot\left[\partial_j\vec{n}\times\partial_k\vec{n}\right]$; here, the magenta surface is defined by $|\vec{\Omega}|=0.5\Omega_{\text{max}}$. Panel (d) shows cylindrical components of $\vec{\Omega}$ in units of $\lambda_h^{-2}$.
		}\label{fig:Sk_vs_sd}
	\end{figure*}
	
	\subsection{Comparison of the screw dislocation sd$_{1}^{\text{Sk}+}$ and the skyrmion tube Sk$^+$ embedded in the conical helix}
	
	The aim of this section is to compare the core structures of the screw dislocation sd$_{1}^{\text{Sk}+}$ and the skyrmion tube Sk$^+$. Furthermore, we will illustrate the merger of both as the negative critical field $H \to - H_{c2}$ is approached from above. 
	
	The key difference between the two configurations is the different behavior of their far field. In the limit $\rho \to \infty$ far away from the core, the magnetization within each $z$-plane is uniform and vortex-like for the skyrmion tube and the screw dislocation sd$_{1}^{\text{Sk}+}$, respectively. This is illustrated in Fig.~\ref{fig:Sk_vs_sd}. As the far field of the skyrmion tube Sk$^+$ is uniform the topological skyrmion number evaluated for each $z$-plane is an integer 
	$N_{\text{top}}=1$ that is independent of the applied magnetic field, see Fig.~\ref{fig:Q-vs-b}. For details on the field evolution of the skyrmion tube within the conical state see Ref.~\cite{Leonov21}.
	
	In contrast, in zero magnetic field $H=0$ the screw dislocation sd$_{1}^{\text{Sk}+}$ forms a meron within each $z$-plane with the skyrmion number $N_{\rm top}=1/2$. This is consistent with the formula $N_{\rm top}=p\nu/2$ for vortices in  two-dimensional ferromagnets \cite{Kosevich90} where $\nu$ is the strength and $p=n_z(\rho=0)$ is the polarity, here $p=1$. In the presence of a finite magnetic field, the mapping of the order parameter from any $z$-plane covers more or less than half of the Bloch sphere for negative and positive $H$, respectively. This is reflected by the field-dependent skyrmion number, $N_{\rm top}=(1-H/H_{c2})/2$ for the screw dislocation, see Fig.~\ref{fig:Q-vs-b}. 
	
	\begin{figure*}
		\includegraphics[width=\textwidth]{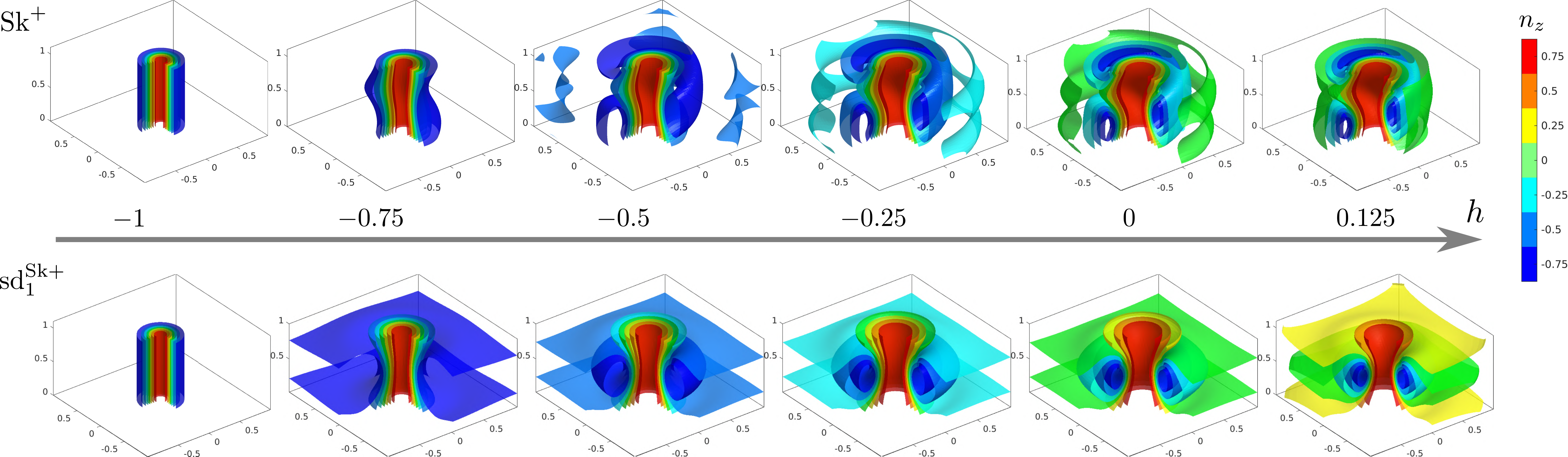}
		\caption{Evolution of $\text{sd}_{1}^{\text{Sk}+}$ and Sk$^+$ in a magnetic field. As the negative critical field is approach $h \to -1$ the two merge into the same skyrmion tube within the field-polarized state. 
			Data is obtained by means of micromagnetic simulations for a cylinder-shaped sample with radius $R_0=5\lambda_h$. Color of the isosurfaces represents the value of the $z$-component $n_z$, see legend. In the insets, the positions on the axes are given in units of $\lambda_h$.
		}
		\label{fig:fld_evol}
	\end{figure*}

	As the negative critical field is approached $H \to - H_{c2}$ the screw dislocation sd$_{1}^{\text{Sk}+}$ and the skyrmion tube Sk$^+$ merge into the same structure, i.e., a skyrmion tube within the field-polarized state, see the illustration in Fig.~\ref{fig:fld_evol}. As discussed in the main text, sd$_{1}^{\text{Sk}+}$ can be considered close to $- H_{c2}$ as a bound state of a skyrmion tube with a vortex of the XY-order parameter of the magnon condensation transition at $H_{c2}$. The latter endows sd$_{1}^{\text{Sk}+}$ with the vortex-like far field. 
	
	The smooth evolution of sd$_{1}^{\text{Sk}+}$ as a function of field $H \to -H_{c2}$ or, equivalently, of sd$_{1}^{\text{Sk}-}$ for $H \to H_{c2}$ is to be contrasted with sd$^+_{-1}$ that switches its core magnetization before reaching the critical field $H_{c2}$, see Fig. 3 of the main text. The latter behavior is similar to vortices in two-dimensional ferromagnets that can also show a switching of their core polarity \cite{Kravchuk07a}.

	\subsection{Screw dislocation sd$^{\rm Bp}_1$ and the energy of its Bloch points}
	
	The aim of this section is to provide some quantitative explanations about the equilibrium positions of the Bloch points (BP) along the core of sd$_1^{\text{Bp}}$ at zero field. Moreover, we discuss the question as to why antivortex-like dislocations sd$_{-1}^\pm$ do not possess BPs.
	
	Let us consider a BP  $\vec{n}_{\textsc{bp}}=c\,\mathfrak{R}\,\vec{r}/|\vec{r}|$ of charge $c$ with $\mathfrak{R}$ being the matrix of spatial rotations \cite{Malozemoff79}. The Bloch point charge $c=\pm1$ is defined by the relation \cite{Malozemoff79}
	\begin{equation}\label{eq:Top-charge}
	\vec{\nabla}\cdot\vec{\Omega}=c\delta(\vec{r}-\vec{r}_{\textsc{bp}}),\qquad \vec{\Omega}=\hat{\vec{x}}_i\frac{\varepsilon_{ijk}}{8\pi}\vec{n}\cdot[\partial_j\vec{n}\times\partial_k\vec{n}],
	\end{equation} 
	where ${\vec{r}_\textsc{bp}}$ denotes the Bloch point position. If $\mathfrak{R}=\mathfrak{R}_{\hat{\vec{z}}}(\alpha)$ describes the rotation by a constant angle $\alpha$ around the $z$-axis, then the BP has a vortex structure within the $(x,y)$-plane similarly to the BPs shown in Fig.~4(b) of the main text. In zero magnetic field, the exchange energy density of the BP  $\mathscr{E}_{\text{ex}}=A(\partial_i\vec{n})^2=2A/r^2$ and the DMI energy density is  $\mathscr{E}_{\textsc{dmi}}=D\,\vec{n}\left[\vec{\nabla}\times\vec{n}\right]=2D\sin\alpha\frac{z}{r^2}$. The DMI contribution vanishes for the integrated energy $E_{\textsc{bp}}$ of a sphere with radius $R_{\textsc{bp}}$ centered on the BP. Finally, we reproduce the well known \cite{Doering68,Pylypovskyi12} result for the BP energy $E_{\textsc{bp}}^0=8\pi A R_{\textsc{bp}}$. 
	
	An important distinction in the present case is that the angle $\alpha$ is not a constant because the BP is immersed into the helical state, namely $\alpha=\frac{2\pi}{\lambda_h}(z-z_{\textsc{bp}})+\frac{\pi}{2}(1-c)$, where $z_{\textsc{bp}}$ is the BP position. The last summand appears by demanding that at $z=0$ the asymptotic value $\vec{n}(\rho\to\infty,\chi=0,z=0)=\hat{\vec{x}}$. For such a position dependent angle $\alpha(z)$, the BP energy has the form
	\begin{equation}\label{eq:EPB}
	E_{\textsc{bp}}=E_{\textsc{bp}}^0\left[1-\left(\frac{2\pi}{3}\right)^2\frac{R_{\textsc{bp}}^2}{\lambda_h^2}+c\,\mathcal{A}\cos\left(2\pi\frac{z_{\textsc{bp}}}{\lambda_h}\right)\right],
	\end{equation}
	with $\mathcal{A}=\frac{\lambda_h}{\pi R_{\textsc{bp}}} \left[\text{Si}(2\pi R_{\textsc{bp}}/\lambda_h)-\sin(2\pi R_{\textsc{bp}}/\lambda_h)\right]\approx2\left(\frac{2\pi}{3}\right)^2R_{\textsc{bp}}^2/\lambda_h^2$ where  $\text{Si}(x)$ is the sine integral and the approximation is valid for $R_{\textsc{bp}} \ll \lambda_h$. Minimization of $E_{\textsc{bp}}$ with respect to $z_{\textsc{bp}}$ results in the equilibrium BP positions
	\begin{equation}\label{eq:BP-Pos}
	z_{\textsc{bp}}^c=\lambda_h\left[\frac{c+1}{4}+ m\right],\qquad m\in\mathbb{Z}.
	\end{equation} 
	These equilibrium BP positions coincide with the ones obtained by means of  micromagnetic simulations for sd$^{\rm Bp}_1$, see Fig.~4(b) of the main text. The resulting BP energy at equilibrium for $R_{\textsc{bp}} \ll \lambda_h$ is then given by $E_{\textsc{bp}}\approx E_{\textsc{bp}}^0\left[1-\frac{(2\pi)^2}{3}R_{\textsc{bp}}^2/\lambda_h^2\right]$.
	
	Let us now estimate the energy of a BP which has an anti-vortex structure in its horizontal cross-section, i.e. $\vec{n}_{\textsc{bp}} =c\,\mathfrak{R}_{\hat{\vec{z}}}(\alpha)\mathfrak{R}_{\hat{\vec{x}}}(\pi)\vec{r}/|\vec{r}|$. A direct calculation shows that the corresponding Bloch point energy  $E_{\textsc{bp}}^\textsc{av}=E_{\textsc{bp}}^0\left[1-\left(\frac{2\pi}{3}\right)^2\frac{R_{\textsc{bp}}^2}{\lambda_h^2}\right]$ is independent on the BP position, and it is higher than the energy of the BP with the vortex-like horizontal cross-section. The latter supports the numerically obtained fact that the anti-vortex screw dislocations sd$_{-1}^\pm$ do not contain Bloch points.
	
	We note that the statics and dynamics of a single BP with a vortex cross-section in a cubic helimagnet was recently studied numerically \cite{Charilaou20} as well as experimentally \cite{Li20b}. 
	
	
\end{document}